\documentclass[reprint,amsfonts, amssymb, amsmath,  showkeys, nofootinbib,pra, superscriptaddress, twocolumn,longbibliography]{revtex4-2}

\usepackage{graphicx}
\usepackage{dcolumn}
\usepackage{bm}
\usepackage{physics}
\usepackage{amsmath}
\usepackage{amsthm}
\usepackage{hyperref}
\usepackage{multirow}
\usepackage{makecell}
\usepackage{bbold}

\usepackage{algorithm}
\usepackage{algorithmicx}
\usepackage{algpseudocode} 

\usepackage{enumerate}

\newtheorem{theorem}{Theorem}
\newtheorem{corollary}{Corollary}
\newtheorem{lemma}[theorem]{Lemma}
\newtheorem{definition}{Definition}

\newtheorem{proposition}{Proposition}

\graphicspath{{Images/}}
\usepackage{xcolor}

\newcommand{\be}{\begin{equation}}
\newcommand{\ee}{\end{equation}}
\newcommand{\XC}{\mathcal{X}}

\newcommand{\LC}{\mathcal{L}}
\newcommand{\CC}{\mathcal{C}}
\newcommand{\AC}{\mathcal{A}}
\newcommand{\HC}{\mathcal{H}}
\newcommand{\OC}{\mathcal{O}}

\renewcommand{\vec}[1]{\boldsymbol{#1}}

\begin{document}

\preprint{APS/123-QED}

\title{A Unified Framework for Trace-induced Quantum Kernels}

\author{Beng Yee Gan}
  \email{gan.bengyee@u.nus.edu}
  \affiliation{Centre for Quantum Technologies, National University of Singapore, 3 Science Drive 2, Singapore 117543}
 
\author{Daniel Leykam}
  \affiliation{Centre for Quantum Technologies, National University of Singapore, 3 Science Drive 2, Singapore 117543}

\author{Supanut Thanasilp}
  \affiliation{Centre for Quantum Technologies, National University of Singapore, 3 Science Drive 2, Singapore 117543}
  \affiliation{Institute of Physics, Ecole Polytechnique Fédérale de Lausanne (EPFL), CH-1015 Lausanne, Switzerland}
  \affiliation{Chula Intelligent and Complex Systems, Department of Physics, Faculty of Science, Chulalongkorn University, Bangkok 10330, Thailand}
  
\date{\today}

\begin{abstract}
Quantum kernel methods are promising candidates for achieving a practical quantum advantage for certain machine learning tasks. Similar to classical machine learning, an exact form of a quantum kernel is expected to have a great impact on the model performance. In this work we combine all trace-induced quantum kernels, including the commonly-used global fidelity and local projected quantum kernels, into a common framework. We show how generalized trace-induced quantum kernels can be constructed as combinations of the fundamental building blocks we coin ``Lego'' kernels, which impose an inductive bias on the resulting quantum models. We relate the expressive power and generalization ability to the number of non-zero weight Lego kernels and propose a systematic approach to increase the complexity of a quantum kernel model, leading to a new form of the local projected kernels that require fewer quantum resources in terms of the number of quantum gates and measurement shots. We show numerically that models based on local projected kernels can achieve comparable performance to the global fidelity quantum kernel. Our work unifies existing quantum kernels and provides a systematic framework to compare their properties.

\end{abstract}

\maketitle
\section{Introduction}

The rise of quantum computers has expanded the realm of data analysis, leading to the fast-evolving field of quantum machine learning (QML)~\cite{biamonte2017quantum}. From the theoretical standpoint, QML aims to understand the fundamental limitations and opportunities of how quantum and classical data can be analyzed using quantum systems. On the other hand, a practical goal of QML is to achieve quantum advantages on some real-world problems. 

Firm understanding of foundational aspects of QML is necessary to develop algorithms with practical advantages over classical machine learning either in computational or sample complexities ~\cite{cerezo2022challenges,schuld2022quantum}. This is particularly relevant in the current era of noisy intermediate-scale quantum (NISQ) devices~\cite{preskill2018quantum, bharti2022noisy} that can only support hundreds of qubits and operations, far below the scale of billions of parameters and data points used in classical machine learning models such as deep neural networks, which makes empirical comparisons infeasible.

Among promising QML algorithms, quantum kernel methods are attracting a great deal of attention due to the well-grounded theoretical tools inherited from the classical kernel theory \cite{scholkopf2002learning,hofmann2008kernel,steinwart2008support,mohri2018foundations}. Here, classical input data points are mapped into quantum states in an exponentially large (in the number of qubits) Hilbert space through the data embedding (also called a quantum feature map). In the case of quantum data, these states are already given. A kernel function that captures the similarity between pairs of states is then measured using the quantum device and the model prediction is obtained from simple classical post-processing of the measurement results. Quantum kernel machines have been rigorously shown to achieve advantages over their classical counterparts on certain \textit{artificial} datasets~\cite{liu2021rigorous,jager2023universal,huang2021power}, and applications to a wide range of scientific and industrial areas such as cosmology~\cite{peters2021machine}, quantum many-body physics~\cite{sancho2022quantum} and finance~\cite{kyriienko2022unsupervised} have also been proposed.

The choice of kernel function has a big impact on the performance of classical machine learning models~\cite{NIPS2012_dbe272ba,duvenaud2014automatic}. The choice of quantum kernel is expected to play a similarly important role. Two popular choices of quantum kernels are (i) global fidelity quantum kernels (GFQKs) \cite{schuld2019quantum,havlivcek2019supervised} and (ii) projected quantum kernels \cite{huang2021power}. As the name suggests, the GFQK is simply a quantum fidelity between states, making a global comparison in the quantum Hilbert space. The projected quantum kernels, on the other hand, use estimations of different local quantities to collectively measure the similarity between quantum states. One common class of the projected quantum kernels is linear projected quantum kernels (LPQKs). 

There have already been theoretical and heuristic studies of quantum kernels on several aspects. For example, the inductive bias of LPQKs (for one subsystem) can be analyzed through spectral decomposition~\cite{kubler2021inductive}. Ref.~\cite{huang2021power} rigorously studied a relative performance between quantum and classical kernel-based models by comparing their generalization bounds. The method for encoding data into quantum states must be chosen with care due to the infamous problem of exponential concentration of kernel values, which leads to poor model performance~\cite{huang2021power, kubler2021inductive, thanasilp2022exponential}. Quantum kernels have also been shown to have close connections to other QML candidates~\cite{schuld2021quantum, jerbi2023quantum}. Furthermore, Ref.~\cite{GilFuster2023expressivity} has recently shown that any valid quantum kernel can be cast as an inner product between two quantum states. Examples of other aspects studied include the optimization of the data embeddings~\cite{hubregtsen2022training} and the role of hyperparameters~\cite{shaydulin2022importance,canatar2022bandwidth}.

Among the GFQK and LPQKs proposed in literature, what is the best way to choose a kernel for a specific learning problem?
A better understanding of the fundamental relations between GFQKs and LPQKs in terms of their expressive power and generalization ability is needed to make an informed kernel selection. In this work, we present a unified framework for generalized trace-induced quantum kernels (GTQKs) that encompasses GFQKs and LPQKs as subsets and reveals the deep connection between GFQKs and LPQKs in terms of their expressive power and generalization ability. We identify the smallest unit of kernels for this class, coined Lego quantum kernels, which we show induce an expressivity structure that enables the comparison of expressive power between different kernels in the class. 

For a system of $n$ qubits there are $4^n$ Lego quantum kernels and the GTQK is defined as the positive weighted linear combinations of these basic kernels. In other words, different quantum kernels are realized by choosing different sets of weights. We illustrate the operational role of the weights through the lens of classical kernel theory, showing how they impose inductive bias on the associated quantum models. Projection into smaller subspaces and composition of subspaces impose an inductive bias on the models, affecting the performance of quantum models associated with LPQKs. Next, we show that the hypothesis class associated with the GTQK resembles the hypothesis class for the multiple kernel learning problem \cite{cortes2010generalization,kloft2011lp,gonen2011multiple}. We use tools from multiple kernel learning to quantify the generalization ability of this family of trace-induced quantum kernels. This allow us to identify the number of Lego quantum kernels as the complexity measure that simultaneously controls the expressivity and generalization error of this class of model. 

Finally, we demonstrate the practicalities of the formalism in the Pauli basis. We propose a natural model selection that systematically increases the complexity of the model, leading to $H$-body LPQKs (where $H$ represent the maximum support of the Pauli operators). Fewer quantum resources are required to implement $H$-body LPQKs compared to the more commonly-used GFQK. These resources include (i) the number of measurement shots (for a relative large training data and fixed $H$) and (ii) the number of quantum gates. $H$-LPQKs demand fewer quantum resources as they can be efficiently estimated using classical shadows \cite{huang2021power}, free of the inversion and SWAP tests, and the access to the training data is not required in the prediction phase. Our numerical examples show that the LPQKs can achieve comparable performance to the GFQKs. This provides empirical evidence in favor of using LPQKs rather than the GFQK.

The outline of this paper is as follows.  Sec.~\ref{Sec:Preliminaries-Frameworks} reviews preliminaries and the types of quantum kernels considered in this work. Next, Sec.~\ref{Sec:Unifying-TIQKs} introduces the unifying framework based on GTQKs and discusses their generalization ability. The practicalities of the framework are illustrated in Sec.~\ref{Sec:practicalities} using the example of fashion-mnist classification. Sec.~\ref{Sec:Conclusion} concludes the paper.

\section{Preliminaries}
\label{Sec:Preliminaries-Frameworks} 
We consider supervised learning tasks with input vectors $\boldsymbol{x} \in \mathcal{X}$ and the associated labels $y \in \mathcal{Y}$, related via a target function $g:\mathcal{X} \rightarrow \mathcal{Y}$. Given labelled training data $S := \{\boldsymbol{x}_i, y_i \}_{i=1}^N$ drawn independently and identically distributed (i.i.d) from a distribution $\mathcal{D} := \mathcal{X} \times \mathcal{Y}$, supervised machine learning algorithms aim to approximate the target function by training a parameterized model $f_{\bf a}(\vec{x})$ with trainable parameters $\vec{a}$. This is achieved by minimizing an empirical risk to find optimal parameters 
\begin{align}\label{eq:empirical-risk}
    {\bf a}^{\rm (opt)} := \text{argmin}_{\bf a} \mathcal{L}_{\bf a} (S) \;.
\end{align}
The hope is that after training the model can \textit{generalize} well. That is, the predictions on unseen input data of the trained model agree with the true labels i.e., $f_{\bf{a}^{(opt)}}(\vec{x}) \approx g(\vec{x})$ for $\vec{x} \notin S$. One way to assess the generalization ability of the model is through the generalization bound 
\begin{align}
    \mathcal{L}_{\bf a} - \mathcal{L}_{\bf a} (S) \le \mathcal{B}(N)
\end{align}
where $\mathcal{L}_{\bf a} = \mathbb{E}_S (\mathcal{L}_{\bf a}(S))$ is the true risk and $\mathbb{E}_S(\cdot)$ is the expectation over all possible data points $S \sim \mathcal{D}^N$. The generalization gap $\mathcal{B}(N)$ captures the model complexity and gets smaller with more training data $N$. Learning is successful when both $\mathcal{L}_{\bf a} (S)$ and $\mathcal{B}(N)$ are small.

Achieving good generalization also depends greatly on two other contributing factors: \textit{expressive power} and \textit{trainability}. The model expressivity informs the complexity of the model class and is related to $\mathcal{B}(N)$, while the trainability of a model tells us about the difficulty of optimizing the empirical loss function in Eq.~\eqref{eq:empirical-risk}. 

\subsection{Kernel-based models}

Kernel methods rely on the mapping of input data into a higher-dimensional feature space. 
\begin{definition}[Kernel function]\label{def:kernel}
    Let $(\XC, \mu)$ be a finite measure space. A kernel function (or simply a kernel) maps a pair of input data points $\vec{x}, \vec{x}' \in \XC$ 
    to some real values $k: \XC \times \XC \rightarrow \mathbb{R}$ and has the following properties
    \begin{enumerate}
        \item Symmetric: for all $\vec{x}, \vec{x}' \in \XC$, 
        \begin{align}
            k(\vec{x}, \vec{x}') = k(\vec{x}', \vec{x}).
        \end{align}
        
        \item Positive semi-definite: for all $c(\vec{x}) \in L_2(\XC)$
        \begin{align}
            \iint_{\XC\times \XC} c(\vec{x}) c(\vec{x}') k(\vec{x},\vec{x}') \mu(d \vec{x}) \mu(d \vec{x}') \ge 0,
        \end{align}
        where $L_2(\cdot)$ is the set of square integrable functions.
    \end{enumerate}
\end{definition}
The kernel can be seen as an inner product between two feature vectors in the feature space. Specifically, given two feature vectors $\Phi(\vec{x}), \Phi(\vec{x}') \in \mathbb{R}^m$ where $m$ is the dimension of the feature space (which could be infinite), their inner product is equal to the kernel function
\begin{align}
    k(\vec{x}, \vec{x}') = \langle \Phi(\vec{x}), \Phi(\vec{x}') \rangle.
\end{align}
Crucially, many feature maps can lead to the same kernel function. One well-known construction is to express a feature map using its eigenbasis functions. That is, if the kernel function satisfies the \textit{Mercer's condition}, it can always be expressed in its eigendecomposition form, 
\begin{align} \label{eq:Standard-Mercer-decompose}
    k(\vec{x}, \vec{x}') = \sum_{i=0}^\infty \gamma_i \phi_i(\vec{x})\phi_i(\vec{x'})
\end{align}
with eigenfunctions $\phi_i(\vec{x})$ and eigenvalues $\gamma_i$, and the Mercer feature map can be constructed as
\begin{align}
    \Psi(\vec{x}) = (\psi_1(\vec{x}), \psi_2(\vec{x}), \cdots)^T,
\end{align}
where $\psi_i(\vec{x}) = \sqrt{\gamma_i}\phi_i(\vec{x})$. Note that the eigenvalue equation is defined as
\begin{align}
    \int_\XC k(\vec{x},\vec{x'})\phi_j(\vec{x'}) \mu(d\vec{x'}) = \gamma_j \phi_j(\vec{x}).
\end{align}

One concept that we largely use in this work is the multiple kernel theory. Given a set of kernel functions $\mathcal{K} = \{ k_i(\vec{x}, \vec{x'}) \}_i$, one can construct a new kernel as a positive linear combination of these kernel functions
\begin{align}
    k_{\rm tot}(\vec{x},\vec{x'}) = \sum_{i = 1}^{|\mathcal{K}|} w_i  k_i(\vec{x}, \vec{x'}) \;,
\end{align}
where $w_i$ are some positive weights that respect the normalization condition $\sum_i w^2_i = 1$. More details regarding the multiple kernel theory are in Appendix~\ref{Appendix:multiple-kernel-learning}.

\subsubsection{Expressivity}
The expressivity can be seen as the flexibility of the functional form generated by a learning model. 
That is, the expressivity measures the size of the hypothesis class (i.e., the set of all possible functions by the model). In kernel methods, the model's expressive power can be analyzed through the lens of the Reproducing Kernel Hilbert Space (RKHS). Functional bases in the RKHS are uniquely associated with the kernel function and the hypothesis class of the kernel-based model, $\HC_k$, can be expressed as
\begin{align}\label{eq:rkhs-expressivity}
    \mathcal{H}_k = \left \{ f(\cdot) = \sum_{i=1}^{\infty} \alpha_i k(\cdot, \boldsymbol{x}_i) ~; ~ \|f\|^2_{\mathcal{H}_k} < \infty  \right\}\;,
\end{align}
where $\vec{x}_i \in \XC$, $\alpha_i \in \mathbb{R}$ are real-valued coefficients and $\| f \|_{\HC_k}$ is the norm in RKHS that respects the reproducing property i.e., $f(\vec{x}) = \langle f(\cdot), k(\cdot,\vec{x})  \rangle_{\HC_k}$. 
 
Alternatively, the RKHS can be constructed using the Mercer feature map
\begin{align} \label{eq:rkhs-standard-Mercer}
    \mathcal{H}_{k} = \left \{ f(\cdot) = \sum_{i=1}^{\infty} \tilde{\alpha}_i \psi_{i}(\cdot) ~; ~ \|f\|^2_{\mathcal{H}_k} < \infty \right\}
\end{align}
with $\tilde{\alpha}_i \in \mathbb{R}$. Note that $\mathcal{H}_{k}$ is a linear space in terms of $\Psi(\cdot)$ and the functions in Eq.~\eqref{eq:rkhs-standard-Mercer} can be obtained by substituting Eq.~\eqref{eq:Standard-Mercer-decompose} into Eq.~\eqref{eq:rkhs-expressivity} with $\tilde{\alpha}_i = \sum_{j=1}^\infty \alpha_j \psi_i(\vec{x}_j)$. The kernel still has the reproducing property $f(\boldsymbol{x}) = \langle f(\cdot), k(\cdot, \boldsymbol{x}) \rangle_{\mathcal{H}_{k}}$ and it enforces the orthogonality condition of $\psi_i(\cdot)$, i.e.: $\langle \psi_{i}(\cdot), \psi_{j}(\cdot) \rangle_{\mathcal{H}_{k}} = \delta_{ij}$. Given two arbitrary functions $f(\cdot) = \sum_i \tilde{\alpha}_i \psi_{i}(\cdot)$ and $g(\cdot) = \sum_j \tilde{\beta}_j \psi_{i}(\cdot)$, the inner product in this space is
\begin{align}
    \langle f, g \rangle_{\mathcal{H}_{k}}  &:= \sum_{i,j=1}^{\infty} \tilde{\alpha}_i\tilde{\beta}_j \langle \psi_{i}(\cdot), \psi_{j}(\cdot) \rangle_{\mathcal{H}_{k}} = \sum_{i=1}^{\infty} \tilde{\alpha}_i \tilde{\beta}_i.
\end{align}
A similar procedure can construct the RKHS using the eigenfunctions $\phi_i$ instead of the Mercer feature map\footnote{Note that here we discuss the construction of the RKHS, rather than the construction of a feature map discussed earlier.}. The detailed discussion of this together with the isometric isomorphic mappings is deferred to Appendix.~\ref{Appendix:Mercer-RKHS}.

The kernel type influences the properties and inductive bias of the resulting hypothesis class. For example, Gaussian and Laplacian kernels respect shift invariant symmetry between two data points. In addition, the Gaussian (Laplacian) kernels generate smooth (rigged) functions.

\subsubsection{Trainability}

One notable strength of kernel-based models is their trainability guarantee. Thanks to the representer theorem, for a given training dataset $S$, the optimal kernel-based model is guaranteed to be of the form
\begin{align}\label{eq:model-prediction}
    f_{\vec{a}^{\rm (opt)}}(\vec{x}) = \sum_{i = 1}^{N} a_{i}^{\rm (opt)} k(\vec{x},\vec{x}_i) \;,
\end{align}
where $\vec{a}^{\rm (opt)} = \left(a_1^{\rm (opt)}, ... , a_{N}^{\rm (opt)}\right)$. The number of trainable parameters scales linearly with the number of training data, in contrast to the number of coefficients $\{\alpha_i \}$ in Eq.~\eqref{eq:rkhs-expressivity}, which can be infinite.

If the loss function $\LC_{\vec{a}}$ is properly chosen, then the optimization problem Eq.~\eqref{eq:empirical-risk} becomes convex. Examples include the square loss function in the kernel ridge regression and the hinge loss function in the binary classification with support vector machines.  

\subsubsection{Generalizability}

Theoretical tools such as RKHS allow us to derive generalization bounds for kernel-based models. As a prime example, we consider a binary classification task with the $\mathcal{C}$-margin loss function with $\mathcal{C} \ge 0$
\begin{align}
    \LC_{\vec{a}}^{(\CC)}(S) = \frac{1}{N}\sum_{i=1}^N \Phi_\CC(y_i f_{\vec{a}}(\boldsymbol{x}_i))
\end{align}
where
\begin{align}
    \Phi_\CC(z) = \begin{cases}
        1 & \text{if} \quad z \le 0\\
        1- \frac{z}{\CC} & \text{if} \quad 0 \le z \le \CC\\
        0 & \text{if} \quad \CC \le z
    \end{cases}\;.
\end{align}
The generalization error can be bounded using the Rademacher complexity~\cite{mohri2018foundations}. The Rademacher complexity is a complexity measure that captures the richness of functions in the associated model class by measuring their ability to fit random noise. In other words, the model class can learn more complex functions if they have a higher Rademacher complexity.  
As the precise definition of Rademacher complexity is not necessary for the discussion, we defer further discussion to Appendix.~\ref{Appendix-Sec:Generalization-bounds-introduction}.

By Theorem 5.8 in Ref.~\cite{mohri2018foundations}, with probability at least $1-\delta$, where $\delta \in [0,1]$, the following bounds hold
\begin{align} \label{Eqn:Generalization-error-margin-Radamecher}
     \LC_{\vec{a}}^{(\CC)}&\le \LC_{\vec{a}}^{(\CC)}(S) + \frac{2}{\CC}\hat{\mathfrak{R}}_S (\HC) + 3\sqrt{\frac{\log \frac{2}{\delta}}{2N}} \\ 
    \mathrm{and} \quad 
     \LC_{\vec{a}}^{(\CC)} &\le \LC_{\vec{a}}^{(\CC)}(S) + \frac{2}{\CC}\mathfrak{R}_N (\HC) + \sqrt{\frac{\log \frac{1}{\delta}}{2N}}.
\end{align}
where $\hat{\mathfrak{R}}_S (\HC)$ is the empirical Rademacher complexity for $\mathcal{H}$ estimated using labeled training sample $S$, while $\mathfrak{R}_N(\HC)$ is the Rademacher complexity obtained by averaging $\hat{\mathfrak{R}}_S (\mathcal{H})$ over all possible samples $S$.

\subsection{Quantum kernels}
We now describe the typical pipeline of quantum kernel methods. Here, quantum computers are used to encode classical input data into quantum states. In particular, this data-embedding process involves embedding each individual input data $\vec{x}$ into an $n$-qubit quantum state through a data-dependent unitary $U(\vec{x})$ such that
\begin{align}
    \rho(\vec{x}) = U(\vec{x})\rho_0 U^\dagger(\vec{x})  \;,
\end{align}
where $\rho_0$ is some initial state. These quantum states are generally in much higher dimensions than the original data space, resembling feature vectors in the classical kernel methods. Indeed, a quantum kernel can be defined as an appropriate choice of a similarity measure between two states that respects Definition~\ref{def:kernel}. 

There are multiple choices of quantum kernel functions. In this work, we focus on two commonly used classes of trace-induced quantum kernels: (i) the global fidelity quantum kernels (GFQKs) and (ii) the linear projected quantum kernels (LPQKs). Examples of other types of quantum kernels in the literature include quantum neural tangent kernels \cite{liu2022representation,shirai2021quantum}, quantum path kernels \cite{incudini2022quantum}, quantum Fisher kernels \cite{suzuki2022quantum}, and Gaussian projected quantum kernels \cite{huang2021power}.

\subsubsection{Global fidelity quantum kernels}
For an input pair $\vec{x}$ and $\vec{x'}$, the global fidelity quantum kernels (GFQKs) compares two quantum states at the global level and is defined as
\begin{align}\label{eq:global-kernel}
    k_n(\vec{x},\vec{x'}) = \tr(\rho(\vec{x})\rho(\vec{x'})) \;.
\end{align}

By the representer theorem, the quantum model can be rewritten as \cite{scholkopf2001generalized}
\begin{align} 
    f_{\vec{a}}(\boldsymbol{x}) &= \sum_{i=1}^N a_i \tr(\rho(\boldsymbol{x}_i)\rho(\boldsymbol{x})) = \tr(M(\vec{a}) \rho(\vec{x}))
\end{align}
where $M({\bm a}) = \sum_{i=1}^N a_i \rho(\boldsymbol{x}_i) $ can be interpreted as the optimal measurement for the quantum model. The measurement of $M(\vec{a})$ requires a fault-tolerant quantum computer to implement~\cite{schuld2021supervised}. An alternative is to measure the quantum kernels by the SWAP test or adjoint method and then train ${\bm a}$ classically.

\subsubsection{Linear projected quantum kernels}

A family of LPQKs has been proposed as an alternative quantum kernel which compares quantum states at the level of subsystems~\cite{huang2021power, kubler2021inductive}. The simplest way to construct the LPQK is to project quantum states onto only one subsystem, leading to the ${\bf s}$-LPQK
\begin{align} \label{eq:s-LPQK}
    k_{\bf s}(\boldsymbol{x},{\vec{x'}}) =  \tr_{{\bf s}}(\rho_{\bf s}(\boldsymbol{x})\rho_{\bf s}(\vec{x'}))
\end{align}
where ${\bf s}$ is the label of the qubits in the subsystem and $\rho_{\bf s}(\boldsymbol{x}) = \tr_{\bar{{\bf s}}}(\rho(\boldsymbol{x}))$ is the associated reduced density matrix (RDM) with $\tr_{\bar{{\bf s}}}(\cdot)$ being a trace out of the rest. The ${\bf s}$-LPQK has an associated ${\bf s}$-quantum model
\begin{align}
    f_{\vec{a}}^{(\bf s)}(\boldsymbol{x}) 
    = \sum_{i=1}^N a_i k_{\bf s}(\boldsymbol{x}_i,\vec{x}) = \tr_{{\bf s}}(M_{\bf s}(\vec{a})\rho_{\bf s}(\boldsymbol{x})) 
\end{align}
where $M_{\bf s} = \sum_
{i=1}^N a_i \rho_{\bf s}(\vec{x}_i)$ is the local Hermitian observable on ${\bf s}$-indexed qubits. 

A large amount of information about the quantum state $\rho(\boldsymbol{x})$ is discarded if one considers only one partition of the entire system. A sufficient number of partitions should therefore be used to capture the necessary amount of information for a given task. One could use all subsystems of size $S = |{\bf s}|$ to build the $S$-LPQK. That is, we denote $\mathbb{S}_S = \{{\bf s}_1, {\bf s}_2, \dots, {\bf s}_W \: |\: |{\bf s}_i| = S\}$ as the set of subsets of $S$ qubits from $n$ qubits with $W = \binom{n}{S}$ being the number of all possible $S$-RDM partitions. The $S$-LPQK is defined as an equally-weighted sum of all $\bf s$-LPQKs
\begin{align}
    k_{S}(\boldsymbol{x},\vec{x'}) = \frac{1}{\sqrt{|\mathbb{S}_S|}}\sum_{{\bf s} \in \mathbb{S}_S} k_{\bf s}(\boldsymbol{x},\vec{x'}) \;.
\end{align}
The corresponding $S$-quantum model is also given by
\begin{align}\label{eq:S-LPQK}
    f_{S}(\boldsymbol{x}) = \sum_{i=1}^N \alpha_i k_{S}(\boldsymbol{x}_i,\boldsymbol{x}) = \frac{1}{\sqrt{|\mathbb{S}_S|}}\sum_{{\bf s} \in \mathbb{S}_S} f_{\bf s}(\boldsymbol{x}),
\end{align}
which is just the uniform sum of the ${\bf s}$-quantum models. The sum is performed over all possible $S$ partitions of $n$ qubits, and this can be generalized to a weighted sum over the partitions.

\section{Generalized Trace-induced Quantum Kernels}
\label{Sec:Unifying-TIQKs}

In this section, we present a unified framework based on generalized trace-induced quantum kernels (GTQKs) that includes the GFQKs and LPQKs considered in the literature and investigate its expressive structure, inductive bias and generalizability.

\subsection{Unified framework}
\label{Subsec:GTQK}

Given an arbitrary orthonormal Hermitian basis $\mathcal{A} = \{A_i\}_{i=1}^{4^n}$ (with $\tr(A_i A_j) = \delta_{ij}$), the ``Lego'' quantum kernel associated with an operator $A_i$ is defined as
\begin{align}\label{eq:lego-kernel}
    k_i(\vec{x},\vec{x'}) = \tr(\rho(\vec{x})A_i) \tr(\rho(\vec{x'})A_i) \;.
\end{align}
The Lego kernel is the most fundamental building block that compares two data points encoded into quantum states, $\rho(\vec{x})$ and $\rho(\vec{x'})$, in the direction of $A_i$. Since $k_i(\vec{x},\vec{x'})$ only concerns one direction of the basis $\AC$, it has very limited expressivity. To build up the expressive power, the generalized trace-induced quantum kernel (GTQK) is defined as the linear combination of all possible Lego kernels which is of the form
\begin{align} \label{Eqn:GTQK-expanded}
    k({\bm x},{\bm x'}) &= \sum_{i=1}^{4^n} 2^n w_i \tr(\rho(\boldsymbol{x})A_i)\tr(\rho({\vec{x'}})A_i)\\ \label{Eqn:GTQK-trace-form}
    & = \tr(\tilde{\rho}(\vec{x})\tilde{\rho}(\vec{x'})),
\end{align}
where $w_i$ are positive weights that satisfy the normalization constraint $\sum_{i=1}^{4^n} w_i^2 = 1$, and $\Tilde{\rho}({\bm x}) = \sum_{i=1}^{4^n} \tr(\rho({\bm x})A'_i) A'_i$ with $A'_i = \sqrt[\leftroot{-2}\uproot{2}4]{2^n w_i}A_i$. While Eq.~\eqref{Eqn:GTQK-trace-form} motivates the name of “generalized”, we mostly stick with Eq.~\eqref{Eqn:GTQK-expanded} in the remainder of the work, since it has a clearer operational meaning. A simple calculation\footnote{$\left (\int_{\XC}c(\vec{x})\tr(\rho(\vec{x})A_i) \mu(d\vec{x}) \right )^2 \ge 0$ for Lego kernels, and GTQKs positive semi-positiveness follows by linearity.} shows both GTQKs and Lego kernels satisfy Definition~\ref{def:kernel}.

By appropriately choosing the weights and basis, different traced-induced kernels can be obtained, including the GFQK as well as families of the LPQKs. This is formally captured in the following proposition.
\begin{proposition} [Recovering the existing traced-induced kernels]
The form of GTQK is reduced to the trace-induced kernels in the literature including
\begin{itemize}
    \item the GFQK in Eq.~\eqref{eq:global-kernel} with $w_i = 1/2^n$ (regardless of the basis $\AC$)
    \item ${\bf s}$-LPQK in Eq.~\eqref{eq:s-LPQK} and $S$-LPQK in Eq.~\eqref{eq:S-LPQK} with appropriate choices of weights in the Pauli basis $\AC = \{ P_i/\sqrt{2^n} \}_{i=1}^{4^n}$, with $P_i$ being Pauli operators. 
\end{itemize}

\end{proposition}
The fact that GTQKs encompass existing trace-induced kernels enables us to analyze their key fundamental properties such as expressivity and generalizability under the same unified framework. 

\subsection{Expressivity and inductive bias of GTQKs} \label{subsec:expressivity-bias-GTQKs}

The expressive power of GTQKs can be captured by their associated RKHS 
\begin{align}\label{eq:gtqk-rkhs-expressivity}
    \mathcal{H}_{\rm G} = & \left \{ \sum_{i=1}^{\infty} \alpha_i \sum_{k=1}^{4^n} 2^n w_k \tr(\rho(\vec{x}_i) A_k) \tr(\rho(\cdot) A_k)  \right\}\; \\ \label{eq:gtqk-rkhs-expressivity-linear}
    = &  \left \{ \sum_{k=1}^{4^n} \tilde{\alpha}_k \sqrt{2^n  w_k}  \tr(\rho(\cdot) A_k) \right\},
\end{align}
where $\tilde{\alpha}_k = \left(\sum_{i=1}^{\infty} \alpha_i \sqrt{2^n  w_k}\tr(\rho(\vec{x}_i) A_k) \right)$. As expected by their construction, the more non-zero weights $w_k$ the larger the expressivity of the hypothesis class can be. This motivates the use of the number of non-zero weights (denoted by $p$) as a model complexity for GTQKs. Both GTQKs with $p=4^n$ and the GFQK have identical expressive power. That is, they share the same RKHS since one can always absorb the weight as a part of $\tilde{\alpha}_k$ i.e., $\tilde{\alpha}'_k = 2^n w_k \tilde{\alpha}_k$. This raises the question of how the choice of weights affects the GTQKs. 

Here, we argue that the weights provide an inductive bias to the model when training with some regularization. To see this, consider the loss function 
\begin{align} \label{Eqn:Cost-function}
    \LC_{\vec{\tilde{\alpha}}}(S) = & \sum_{i=1}^{N} \ell (f_{\vec{\tilde{\alpha}}}(\boldsymbol{x}_i),y_i) + \frac{\lambda}{2}\|f\|_{\HC _{\rm G}}^2 \; , 
\end{align}
where $\ell (f_{\vec{\tilde{\alpha}}}(\boldsymbol{x}_i),y_i)$ measures how much $f_{\vec{\tilde{\alpha}}}(\boldsymbol{x}_i)$ agrees with $y_i$ and $\lambda$ is the regularization hyperparameter. Explicitly, $\|f\|^2_{\HC _{\rm G}} = \sum_{i=1}^{4^n} \tilde{\alpha}_k^2$, which means all coefficients $\tilde{\alpha}_k$ are equally suppressed by this regularization. At the same time, the functional components in the direction of $A_i$ are scaled with the weights $w_i$. Consequently, the models with coefficients that are aligned in the same directions as large weights are favoured. Thus, GFQKs are equally sensitive to to all degrees of freedom in the Hilbert space, whereas GTQKs with non-uniform weights will be biased to certain directions $A_i$ with larger $w_i$.

There is one caveat here. In general, despite having an orthornormal basis $\AC$, these directions are not orthornormal in the RKHS i.e., $\int \tr(\rho(\vec{x})A_i)\tr(\rho(\vec{x})A_i) \mu(d\vec{x}) \neq \delta_{ij}$. Nevertheless, having more directions generically implies more expressivity.

\begin{figure*}[tb]
\includegraphics[width=\textwidth]{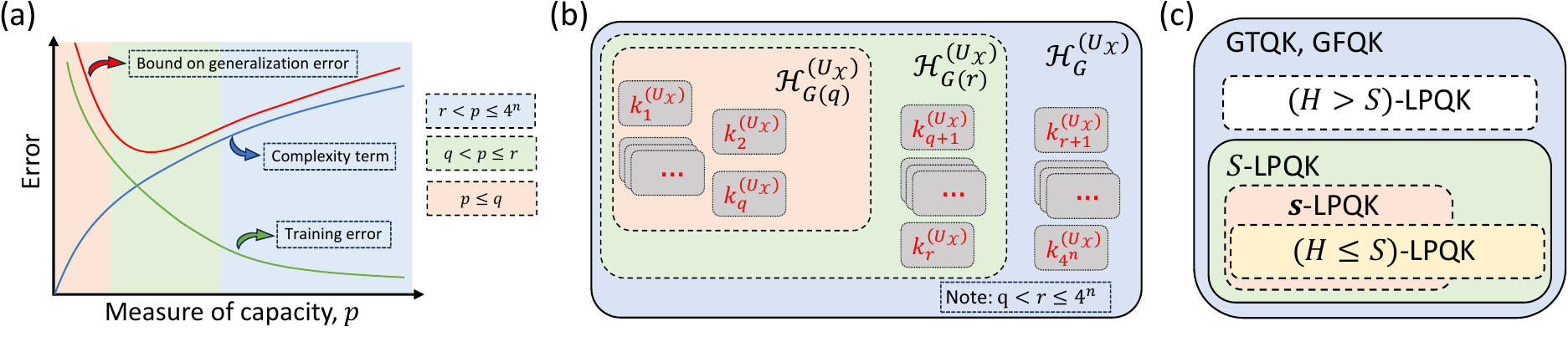}
\caption{\label{Fig:Summary-results} (a) The number of nonzero model weights $p$ simultaneously controls the expressive power and generalization ability of GTQKs, allowing use of structural risk minimization to find the optimal $p$ that minimizes the generalization error. (b) Expressivity hierarchy of nested subsets for trace-induced quantum kernels. (c) GTQKs and GFQKs have the same expressive power, since they contain the same number of Mercer Lego quantum kernels. S-LPQKs consist of Lego quantum kernels generated by up to $S$-body Pauli observables, hence they are more expressive than ${\bf s}$-LPQKs for $|{\bf s}|  = S $ and $H$-LPQK for $H\le S$.}
\end{figure*}

\subsection{Generalizability}
We now discuss the generalization bound with $p$ as the complexity measure. Using the multiple kernel learning theory in Appendix~\ref{Appendix:multiple-kernel-learning} \cite{cortes2010generalization,kloft2011lp}, the empirical Rademacher complexity of GTQKs with $p$ non-zero weights is bounded by
\begin{align}
    \hat{\mathfrak{R}}_S (\HC_{\rm G}) & \le \frac{\sqrt{2\eta_0 ||{\bf u}||_2}}{N} \\ 
    & \le \sqrt{\frac{2\eta_0 \sqrt{p}R^2}{N}} \;,
\end{align}
where ${\bf u} = (\Tr[K_1],\dots,\Tr[K_p])^T$, with $K_i$ being the Gram matrix associated with an $i^{\rm th}$ Lego kernel, $\eta_0 = \frac{23}{22}$, and the second equality is obtained by assuming $K_i(\boldsymbol{x},\boldsymbol{x}) \le R^2$ for all $\boldsymbol{x} \in \mathcal{X}$. Plugging this bound into Eq.~\eqref{Eqn:Generalization-error-margin-Radamecher} yields
Theorem~\ref{Theorem:Generalization-pTIQK}. 

\begin{theorem}[Binary classification margin bound for GTQKs\label{Theorem:Generalization-pTIQK}] Let $\mathcal{H}_{\rm G}$ be the hypothesis class in Eq.~\eqref{eq:gtqk-rkhs-expressivity-linear} corresponding to the GTQKs with $p$ non-zeros weights. For any training dataset $S$ of size $N$ and for any $\delta \in [0,1]$, with probability at least $1-\delta$, the following generalization bound holds for all functions in $\mathcal{H}_{\rm G}$
\begin{align} \label{Eqn:Generalization-bound-pLPQK}
     \LC_{\vec{a}}^{(\CC)} &\le \LC_{\vec{a}}^{(\CC)}(S)  + \frac{2 p^{\frac{1}{4}}}{\CC}\sqrt{\frac{2 \eta_0 R^2}{N}} + 3\sqrt{\frac{\log \frac{2}{\delta}}{2N}},
\end{align}
where $\eta_0 = \frac{23}{22}$. Hence, the generalization error scales as $O(p^{\frac{1}{4}})$.
\end{theorem}
Theorem~\ref{Theorem:Generalization-pTIQK} reveals that $p$ simultaneously controls both the expressivity and the generalization error, hence, one can perform structural risk minimization for this kernel family to obtain an optimal $p$, as illustrated in Fig.~\ref{Fig:Summary-results}(a).

We emphasize that the generalization bound here is derived for the entire hypothesis class of all GTQKs with $p$ non-zero weights. By fixing a set of weights (essentially corresponding to one particular GTQK), the bound can be reduced to the typical kernel bound independent of $p$.

\subsection{Eigenbasis of GTQKs}
So far, we largely leave the choice of the basis $\AC$ arbitrary. In this section, we discuss the eigenbasis (also called the Mercer basis) which provides some theoretical insights into GTQKs.

For a given choice of the data-embedding and input space, the covariance matrix $\int_{\mathcal{X}} \rho(\boldsymbol{x}) \otimes \rho(\boldsymbol{x}) ~\mu(d\boldsymbol{x})$ can be diagonalized to obtain the Mercer basis $\AC_{U_\XC} = \{ A_i^{(U_\XC)}\}_{i=1}^{4^n}$, with associated eigenvalues $\gamma_i$~\cite{kubler2021inductive,canatar2022bandwidth} (also see Appendix~\ref{Appendix-Subsubsec:Eigendecomposition-GFQK} for detailed derivations). A key additional property of this basis is
\begin{align}
    \int_{\XC} \tr(\rho(\vec{x})  A_i^{(U_\XC)}) \tr(\rho(\vec{x})  A_j^{(U_\XC)}) \mu(d\vec{x}) = \sqrt{\gamma_i \gamma_j} \delta_{ij} \;.
\end{align}
Crucially, this leads to different \textit{orthonormal} directions constructed by the Lego kernels; they form orthonormal Lego RKHSs as shown in Proposition ~\ref{Lemma:Orthogonality-Mercer-RKHS}.

\begin{proposition}[Orthogonality between Lego RKHSs]\label{Lemma:Orthogonality-Mercer-RKHS}  Consider Lego kernels $k_j^{(U_\XC)}(\vec{x},\vec{x'}) = \psi_j^{(U_\XC)}(\vec{x})\psi_j^{(U_\XC)}(\vec{x'})$ with feature map $\psi_j^{(U_\XC)}(\boldsymbol{x}) = \tr(\rho(\boldsymbol{x})A_j^{(U_\XC)})$. They have one eigenfunction $\phi_j^{(U_\XC)}(\cdot) = \frac{\tr(\rho(\cdot)A_j^{(U_\XC)})}{\sqrt{\gamma_j}}$ with eigenvalue $\gamma_{j}$, and their associated RKHS is
\begin{align}
    \mathcal{H}_{{\rm G}_j}^{(U_\XC)} = & \left \{ \sum_{i=1}^{\infty} \alpha_i \tr(\rho(\vec{x}_i) A_j^{(U_\XC)}) \tr(\rho(\cdot) A_j^{(U_\XC)})  \right\},
\end{align}
with $\alpha_i \in \mathbb{R}$. The RKHS of $k^{(U_\XC)}_m$ and $k^{(U_\XC)}_n$ will be orthogonal to each other when $m\ne n ~\forall m,n$.
\end{proposition} 

The GTQKs in the Mercer basis can be expressed as a positive linear combination of these Lego kernels
\begin{align}
    k^{(U_\XC)}(\vec{x},\vec{x'}) &= \sum_{k=1}^{4^n} 2^n w_k \tr(\rho(\vec{x}_i) A_k^{(U_\XC)}) \tr(\rho(\vec{x'}) A_k^{(U_\XC)})
    %
\end{align}
with associated eigenfunctions $\phi_i^{(U_\XC)} = \frac{\tr(\rho(\vec{x})  A_i^{(U_\XC)})}{\sqrt{\gamma_i}}$ and eigenvalues $2^n w_k \gamma_k$. Thus, expressing the GTQK in its eigenbasis provides an additional operational interpretation of the role of the weights $w_k$: they re-scale the associated eigenvalues $\gamma_k$ obtained by diagonalizing the covariance matrix (pre-determined by the choice of the data-embedding and input space).

Another interesting consequence is that the RKHS of the GTQK can be decomposed into an internal direct sum of the RKHS of Lego kernels in Mercer basis.
\begin{corollary} [Orthogonal decomposition of the RKHS of GTQK]
By Proposition~\ref{Lemma:Orthogonality-Mercer-RKHS}, the RKHS $\mathcal{H}_G^{(U_\XC)}$ of the GTQK can be orthogonally decomposed into a internal direct sum of RKHS of Lego kernels in Mercer basis, $\mathcal{H}_{G_i}^{(U_\XC)}$
\begin{align}
    \mathcal{H}_{G}^{(U_\XC)} = \bigoplus_{i=1}^{4^n} \mathcal{H}_{G_i}^{(U_\XC)}.
\end{align}
\end{corollary}
This shows that the number of non-zero weights $p$ represents the number of eigenfunctions the GTQKs has access to and hence precisely controls the expressive power. As a consequence, an expressivity hierarchy can be obtained for this kernel class by increasing the number of non-zero weights $q < r < 4^n$, 
\begin{align}
    \HC^{(U_\XC)}_{G(q)} \subset \HC^{(U_\XC)}_{G(r)} \subset \HC_{G}^{(U_\XC)} \; ,
\end{align}
where $\HC^{(U_\XC)}_{G(q)}$ and $\HC^{(U_\XC)}_{G(r)}$ are the RKHS of GTQKs with $q$ and $r$ Mercer Lego kernels respectively, as shown in Fig.~\ref{Fig:Summary-results}(b).

\subsubsection{Mercer LPQKs and their expressvitiy hierarchy}

The hierarchy formed by Mercer Lego kernels prompts a fundamental question as to whether other trace-induced kernels obey a similar structure. To answer this question, we introduce Mercer LPQKs. In particular, we define Mercer ${\bf s}$-LPQKs similar to ${\bf s}$-LPQK in Eq.~\eqref{eq:s-LPQK} as
\begin{align}
    k_{\vec{s}}^{(U_\XC)}(\vec{x},\vec{x'}) = \tr_{\vec{s}} (\sigma_{\vec{s}} (\vec{x})\sigma_{\vec{s}} (\vec{x'})) \; ,
\end{align}
where $\sigma_{\vec{s}} = \tr_{\vec{\bar{s}}}( U_0 \rho(\vec{x}) U^\dagger_0)$, with $U_0$ as a basis transformation matrix that transforms the Mercer basis to the normalized Pauli basis, i.e.: $A_i^{(U_\XC)} = U^\dagger_0 \bar{P}_i U_0$ where $\bar{P}_i \in \mathcal{P}^n = \{\mathbb{1}, X, Y, Z\}^{\otimes n}/\sqrt{2^n}$ is the normalized Pauli basis for an $n$-qubit system. Expanding the Mercer ${\bf s}$-LPQKs in the normalized Pauli basis yields
\begin{align}
    k_{\vec{s}}^{(U_\XC)} (\vec{x},\vec{x'}) &= \sum_{i=1}^{4^S}\tr_{{\bf s}}(\sigma_{\bf s}(\vec{x}) \bar{P}^i_{\bf s}) \tr_{{\bf s}}(\sigma_{\bf s}(\vec{x'}) \bar{P}^i_{\bf s})\\
    &= \sum_{i=1}^{4^S}\tr(\sigma(\vec{x}) \bar{P}^i_{\bf s} \otimes \bar{\mathbb{1}}_{\bar{{\bf s}}}) 
    \tr(\sigma(\vec{x'}) \bar{P}^i_{\bf s} \otimes \bar{\mathbb{1}}_{\bar{{\bf s}}}), 
\end{align}
where the operators $\bar{P}^{i}_{\bf s}$ are the Pauli observables for subsystem ${\bf s}$ of size $S$, i.e.: $\bar{P}^{i}_{\bf s} \in \mathcal{P}^S_{{\bf s}} = \{\mathbb{1}, X, Y, Z\}^{\otimes S}_{{\bf s}}/\sqrt{2^S}$ with $|\mathcal{P}^S_{{\bf s}}| = 4^S$ and $\mathcal{P}^S_{{\bf s}} \otimes \bar{\mathbb{1}}_{\bf \bar{s}} \subset \mathcal{P}^{n}$. The Mercer $S$-LPQK can be defined in the same fashion as Eq.~\eqref{eq:S-LPQK} with $k_{\vec{s}}(\vec{x},\vec{x'})$ replaced by $k_{\vec{s}}^{(U_\XC)}(\vec{x},\vec{x'})$.

In addition, we introduce a new quantum kernel, Mercer $H$-body LPQKs
\begin{align}
    k_H(\boldsymbol{x},{\bf y}) =  \frac{1}{\sqrt{d_H}} \sum_{i = 1}^{d_H}\tr(\sigma(\boldsymbol{x}) P^i_{H}) \tr(\sigma({\bf y}) P^i_{H}),
\end{align}
where $P^i_H$ are $H$-body Pauli observables that act non-trivially only on $H$ qubits,
\begin{align}
    P^i_{H} \in \mathcal{P}_H = \{ \{X, Y, Z\}^{\otimes H}_{{\bf h}}\otimes \mathbb{1}_{{\bf \bar{h}}} \mid |{\bf h}| = H, \forall {\bf h}\}.
\end{align}
The set $\mathcal{P}_H$ contains all $H$-body Pauli observables and has size $|\mathcal{P}_H| = d_H = \binom{n}{H} \cdot 3^H$ \footnote{An example of a $2$-body Pauli observable is $P_1 \otimes P_2 \otimes \mathbb{1}_{3,\dots,n}$ for $P_j \in  \{X,Y,Z\}$ while the $0$-body Pauli observable corresponds to the identity operator.}. One can write $S$-LPQKs in terms of $H$-body LPQKs and vice versa, as shown in Appendix.~\ref{Appendix-Subsubsec:S-LPQK} and ~\ref{Appendix-Subsubsec-HLPQK}, respectively.

So far, what we have done was to define the GTQK and LPQKs in terms of the Mercer Lego kernels. Expressing these kernels in terms of the Mercer Lego kernels enables us to relate the expressivity between existing trace-induced quantum kernels, as shown in Fig.~\ref{Fig:Summary-results}(c). 

\subsubsection{Inductive bias of Mercer LPQKs} \label{Subsubsec:InductiveBias-SLPQK}
Now we will use this framework to reveal how the inductive bias is imposed by projection and summation in the Mercer LPQKs. ${\bf s}$-LPQKs share the same eigenvalues and eigenfunctions with GFQKs, just a smaller set. Hence, projection imposes the inductive bias by removing the eigenfunctions associated with ${\bf \bar{s}}$-subsystems and those arising from higher order correlations, imposing a bias towards the remaining eigenfunctions.

In contrast, the $S$-LPQK takes all partitions into account, not removing any eigenfunctions arising from lower order correlations, i.e: eigenfunctions constructed using $H$-body Pauli observables for $H \le S$. Therefore, the $S$-LPQK only removes the higher order correlation eigenfunctions. Moreover, summing the ${\bf s}$-LPQK induces degeneracy $D^{(S,H)}$ in the $H$-body subspaces, hence, the $S$-quantum model $f_S(\boldsymbol{x})$ is biased towards the eigenfunctions with lower order correlations, since $D^{(S,H)}$ is higher for smaller $H$. In this case, the constant function has the highest contribution as $D^{(S,H)} = \binom{n-H}{S-H}$, i.e.: for $H=0$, $D^{(S,0)} = \frac{n!}{S!(n-S)!}$. This is unfavorable as the learning ability of the corresponding quantum model will likely be low if the constant function dominates. The constant function contribution can be removed by centering the feature map~\cite{kubler2021inductive,heyraud2022noisy}. Our analysis reveals an important reason to center the quantum kernels, especially when one considers the composition of different ${\bf s}$-LPQKs: Summing the ${\bf s}$-LPQKs will induce degeneracy on the local sub-spaces, resulting in quantum kernels dominated by the constant function.

\section{Practicality of the unified framework}
\label{Sec:practicalities}

One limitation of the Mercer basis is that it is in general non-trivial to find, as it is highly dependent on the data embedding as well as the data distribution. Specifically, the data distribution is generally unknown since one only has access to the sampled dataset, not the original data distribution. In addition, the basis is generally hard to find for an arbitrary data embedding even if the data distribution is known.

In this section we investigate the practicality of the GTQKs where the Pauli basis is chosen. Interestingly, the GTQK in this basis still captures GFQK, $s$-LPQKs, $S$-LPQKs, and $H$-body LPQKs, all in Pauli basis, as subsets, preserving the expressivity structure for these trace-induced quantum kernels. In what follows, we will compare the number of measurement shots required to reach a certain accuracy among different GTQKs and the GFQK. The shot scaling with respect to the size of the training dataset is more favourable for GTQKs. In addition, we show numerically how the number of non-zero weights $p$ affects the prediction accuracy. This provides some empirical evidence that good performance comparable to the maximum expressive GFQK can be achieved with some reasonable $p$. Together, this advocates the use of GTQKs with $p$ non-zero weights over the GFQK. 

\subsection{Practical strategy to select GTQKs}

To construct the GTQK in Pauli basis, one first computes the Pauli expectation values $\{ \tr(\rho(\vec{x}) P_i) \}_{i=1}^p$ using a quantum device, and then perform classical post-processing to incorporate the weights and combine the expectation values. It is important to note that a systematic way to choose Pauli observables for the GTQKs is required to orderly build up the complexity of GTQKs. The most natural way is to start off with low body Pauli observables, e.g.: 1-body Pauli observables, and slowly expand the GTQKs by higher body Pauli observables. Without loss of generality, we compare the measurement shot scaling between GFQKs and $H$-body LPQKs, since the shot scaling for other LPQKs can be obtained in a similar fashion.

\subsection{Measurement shot scaling}

We first study the scaling of the measurement shots of the $H$-body LPQKs and compare with the GFQK. Importantly, we find that, for a large number of training data, the measurement shots scaling is in favour of the $H$-body LPQKs for some fixed $H$, with a linear scaling for $H$-body LPQKs and a quadratic scaling for the GFQK. Although we select to study this particular version of LPQKs, this result generally applies for any GTQKs with $p$ non-zero weights in low body Pauli operators.

\begin{figure}[tb]
\includegraphics[width=80mm,scale=0.5]{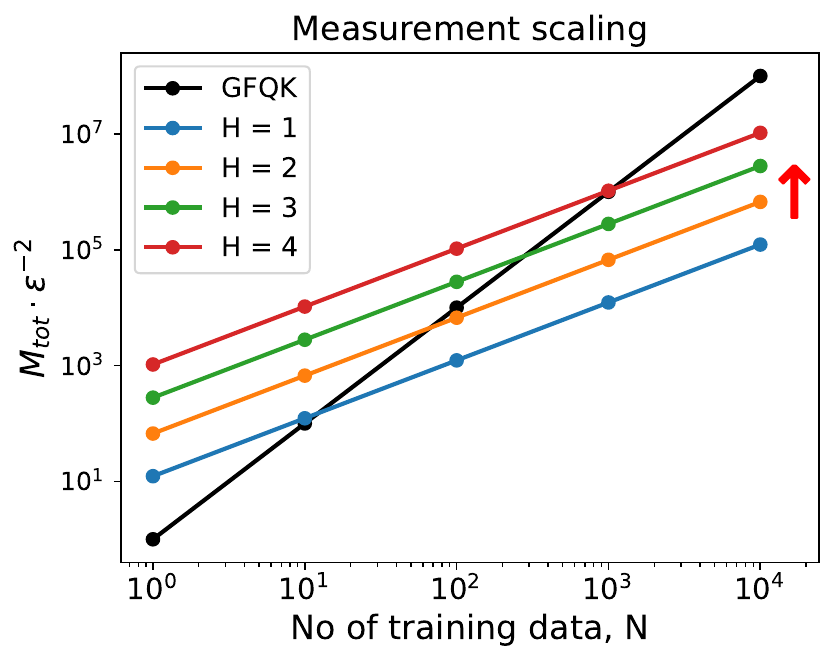}
\caption{The total number of measurements $M_{tot}$ required to estimate $H$-body LPQKs and GFQK for arbitrary approximation error $\varepsilon$ of the kernel matrix elements for a given number of training data points $N$. We consider here a 20-qubit system, but the trend is consistent across different numbers of qubits: The number of measurements grows more rapidly for the GTQK than the $H$-body LPQKs with the number of training data $N$. These scaling behaviors are valid for all $\varepsilon$, since both quantum kernels have the same $\varepsilon$ dependency. 
\label{Fig:Measurement-scaling} }
\end{figure}

Specifically, the GFQK requires $O(\frac{N^2}{\varepsilon^2})$ measurements for all $O(N^2)$ pairwise combinations and arbitrary approximation error $\varepsilon$. On the other hand, we can apply classical shadows to estimate $H$-body LPQKs, leading to shot scaling of $O\big(\frac{\log(|\mathcal{P}_H|)3^H}{\varepsilon^2}N \big)$ for all $O(N)$ data points for the whole $\mathcal{P}_H$ with $|\mathcal{P}_H|=\binom{n}{H} \cdot 3^H$. Fig.~\ref{Fig:Measurement-scaling} compares the total number of measurement shots required to estimate the kernel matrix elements of GFQKs and $H$-body LPQKs for a given number of training data points and with arbitrary approximation error $\varepsilon$. A direct comparison shows that $H$-body LPQKs require fewer measurement shots than the GFQKs when $N > \log(|\mathcal{P}_H|)3^H$. Their measurement costs can be further reduced by using derandomized classical shadows~\cite{huang2021efficient}.

\begin{figure}[tb]
\includegraphics[width=80mm,scale=0.5]{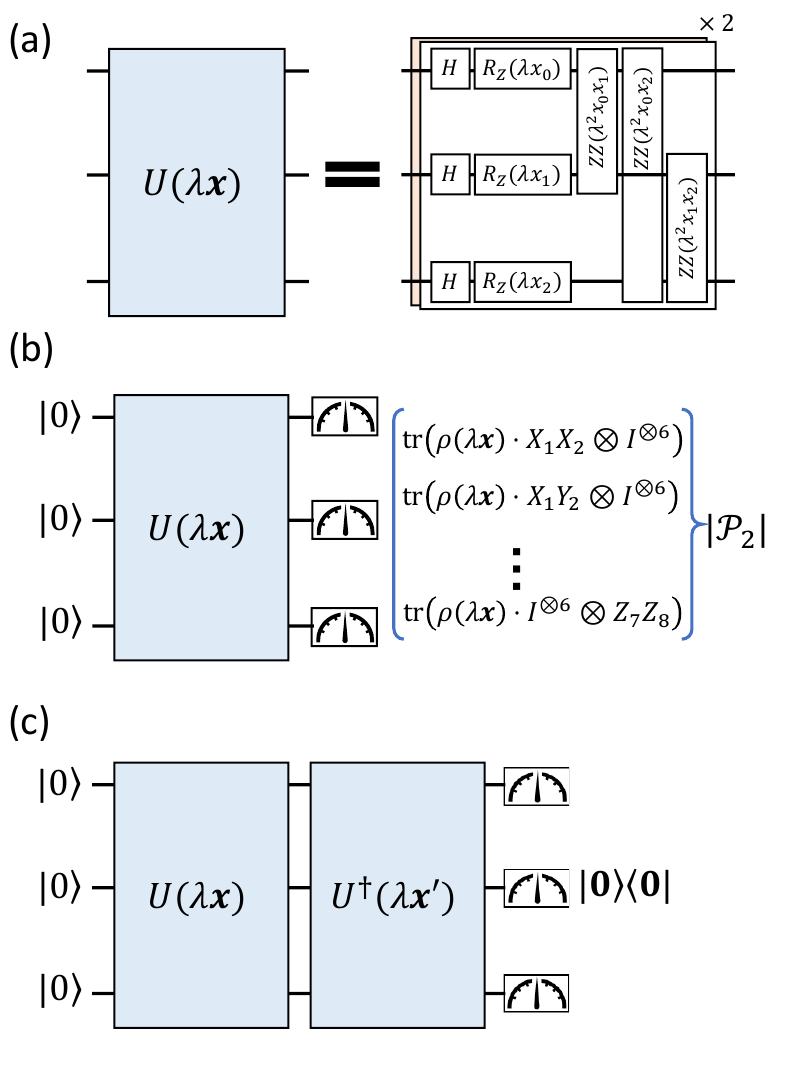}
\caption{(a) The instantaneous quantum polynomial ansatz for a 3-qubit system \cite{havlivcek2019supervised}. (b) The measurement protocol to obtain $2$-body LPQKs. Estimation of $2$-body LPQKs only requires calculation of expectation values of the feature map $\rho(\lambda \boldsymbol{x})$ with respect to 2-body Pauli operators $P^i_2 \in \mathcal{P}_2$. They can be efficiently estimated with random Pauli classical shadows \cite{huang2020predicting} or derandomized classical shadows \cite{huang2021efficient}. (c) The inversion test to estimate the GFQK. In contrast to (b), it requires two copies of $U(\cdot)$, resulting in a deeper quantum circuit. Evaluating the GFQK using the SWAP test reduces the circuit depth but doubles the number of qubits required~\cite{hubregtsen2022training}. \label{Fig:Circuit-schemetics}}
\end{figure}

There are other advantages in favor of $H$-body LPQKs over the GFQK. We summarize all the advantages here:
\begin{enumerate}[{(1)}]
    \item GFQK requires $O(N^2)$ circuit runs for pairwise evaluations, while LPQKs only need $O(N)$ runs for $N$ training data for fixed $H$.
    \item $H$-body LPQKs require shallower circuits that implement only $U(\boldsymbol{x})$ (Fig.~\ref{Fig:Circuit-schemetics}(b)), while the GFQK requires deeper circuits to implement $U(\boldsymbol{x})U^\dagger(\boldsymbol{x}')$ (Fig.~\ref{Fig:Circuit-schemetics}(c)).
    \item  Access to training data in the prediction phase is not required for LPQKs as LPQKs evaluate each data point separably, while the GFQK require pairwise evaluations of data points.
    \item When $N \gg p$, one can solve the feature map form of the model, i.e.: $f(\boldsymbol{x}) = {\bf a}' \cdot {\bm \psi}(\boldsymbol{x})$ instead of the kernel form of the function $f(\boldsymbol{x}) = \sum_{i=1}^{N} a_i k(\boldsymbol{x}_i,\boldsymbol{x})$ for LPQKs with associated feature maps ${\bm \psi}(\cdot)$. One possible application of this is to solve a primal problem instead of a dual problem in support vector machines.
\end{enumerate}

\subsection{Empirical study of model performance with $H$-body LPQKs}

Prediction accuracy on testing data points is one of the deciding factors to choose one kernel over the others. In this section, we numerically show that the LPQKs with a reasonable number of Lego kernels have competitive performance compared to the model with GFQKs. This suggests the use of LPQKs since they consume less resources while maintaining a competitive prediction accuracy.
 
We consider an 8-qubit model with the circuit ansatz $U(\lambda \boldsymbol{x})$ depicted in Fig.~\ref{Fig:Circuit-schemetics}(a) to perform binary classification on the fashion-minist dataset \cite{xiao2017fashion}. The data is standardized and downscaled to 8 dimensions using principle component analysis. The instantaneous quantum polynomial unitary \cite{havlivcek2019supervised} $U(\lambda\boldsymbol{x})$ with kernel bandwidth $\lambda \in [0,1]$ \cite{shaydulin2022importance} is then used to embed the downscaled data points $\boldsymbol{x}$ into the feature space via $\rho(\lambda\boldsymbol{x}) = U(\lambda\boldsymbol{x})\rho_{0}U^\dagger(\lambda\boldsymbol{x})$ for the initial state $\rho_0$. The kernel bandwidth $\gamma$ controls the size of the model's explorable feature space, hence serving as a regularization parameter that limits the expressivity of the model. We construct two types of kernels using $\rho(\lambda\boldsymbol{x})$: (1) $2$-body LPQK with $p$ non-zero weights,
\begin{align}
    k_2^p(\lambda\boldsymbol{x},\lambda{\bf y}) = \sum_{i=1}^p \frac{1}{\sqrt{p}} \tr(\rho(\lambda\boldsymbol{x})P^i_2)\tr(\rho(\lambda{\bf y})P^i_2),
\end{align}
where $P^i_2 \in \mathcal{P}_2 = \{ \{X,Y,Z\}^{\otimes 2}_{\bf h} \otimes \mathbb{1}_{\bar{\bf h}}$ $\forall|{\bf h}| = 2 \}$ are 2-body Pauli operators that act on qubits of index ${\bf h}$ and $p=\big[1,|\mathcal{P}_2| \big]$. For 8 qubits, $|\mathcal{P}_2| = 252$ and we used random Pauli classical shadows \cite{huang2020predicting} to estimate $\tr(\rho(\lambda\boldsymbol{x})P^i_2)$. (2) GFQK: $k_{n}(\lambda\boldsymbol{x},\lambda{\bf y})=\tr(\rho(\lambda\boldsymbol{x})\rho(\lambda{\bf y}))$, estimated using the inversion test \cite{hubregtsen2022training}. The support vector machines with the corresponding kernels are then optimized to perform binary classification on the downscaled dataset. Both GFQK and LPQKs are extracted using Pennylane \cite{bergholm2018pennylane}, while the SVC module in the scikit-learn package \cite{scikit-learn} is used to perform the classification task. All numerical examples use 100 shots per GFQK matrix element and 4000 classical shadows to estimate $2$-LPQKs.

\begin{figure}[tb]
\includegraphics[width=80mm,scale=0.5]{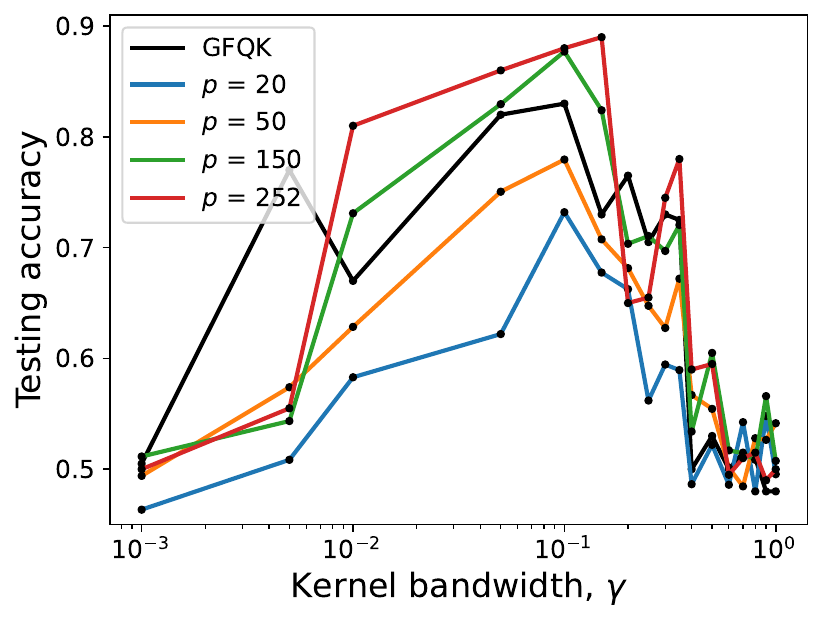}
\caption{Average classification testing accuracy of LPQKs with $p=20,50,150,252$ and GFQK for different kernel bandwidths $\gamma$ with optimized regularization parameter $C^* \in \{0.006,0.015, 0.03, 0.0625, 0.125, 0.25, 0.5, 1.0, 2.0, 5.0, 8.0, 16.0,$ $32.0, 64.0, 128.0, 256, 512, 1024\}$ for 100 training and 20 test data points. For intermediate kernel bandwidths, the accuracy of the 2-body LPQK models is comparable to the GFQK for sufficiently large $p$. \label{Fig:Kernel-bandwidth-results}}
\end{figure}

Fig.~\ref{Fig:Kernel-bandwidth-results} shows the average prediction accuracy of $2$-LPQKs with $p=20,50,150,252$ (nested subsets) and the GFQK for different kernel bandwidths $\gamma$ and the optimized regularization parameter $C^*$. This parameter $C^*$ is obtained using 10-fold cross-validation and the results for $p$-LPQKs are averaged over 10 different random selections of $p$ features. The LPQKs perform equally bad as the GFQK for the small and large kernel bandwidth regimes, indicating that LPQKs cannot be used to resolve the issues faced by the GFQK. For intermediate kernel bandwidth regimes, the expressive power and classification accuracy of 2-LPQK models increase with $p$. The 2-LPQK achieves comparable accuracy to the GFQK for sufficiently large $p$, showing the capabilities of LPQKs for classification tasks.

\begin{figure}[tb]
\includegraphics[width=80mm,scale=0.5]{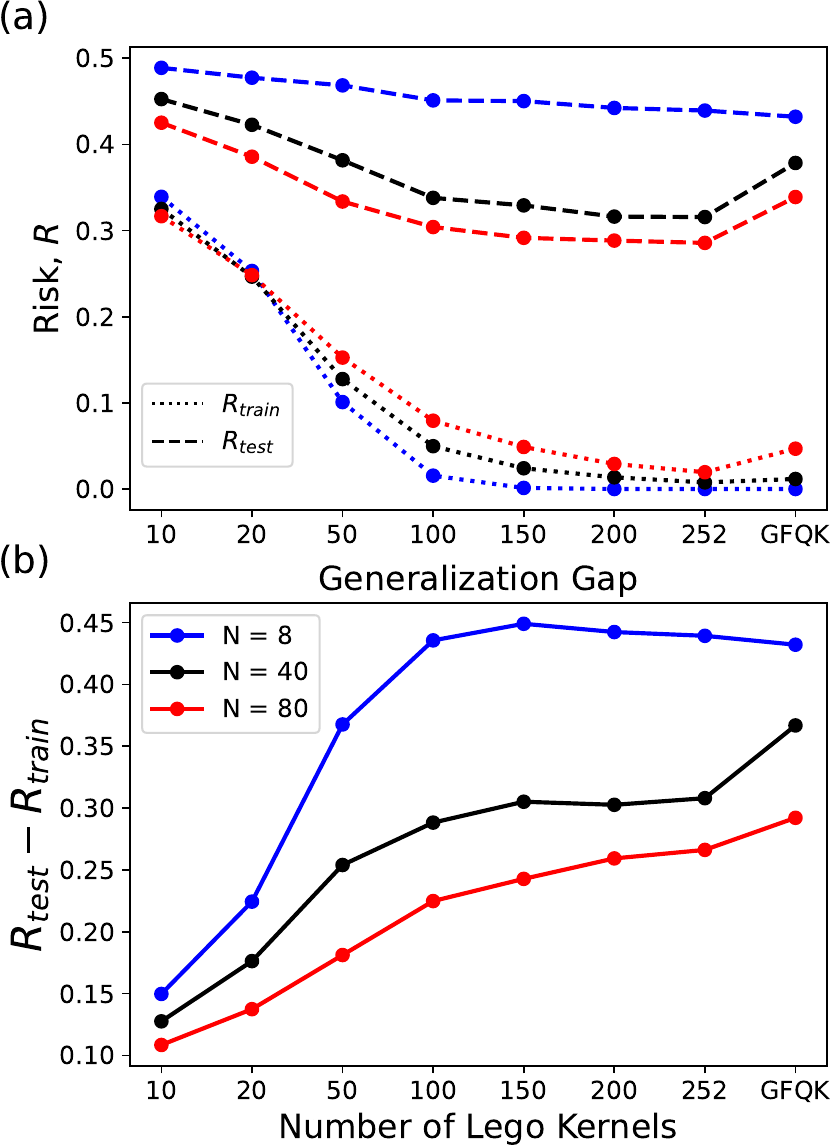}
\caption{(a) The training and testing risk (Risk = 1 - Accuracy) and (b) the empirical generalization gap between training and testing risk of models associated to the GFQK and $2$-body LPQKs with $p=10,20,50,100,150,200,252$ for $\gamma = 0.2$, $C = 5$ and $N = 8, 40, 80$. Increasing $p$ simultaneously increases the training classification accuracy and the expressivity of the model, resulting in an increasing generalization gap for the LPQKs models. However, a highly expressive model will overfit the training data, as illustrated in the increased testing risk of GFQK in (a) for $N=40$.
\label{Fig:Generalization_error} }
\end{figure}

Fig.~\ref{Fig:Generalization_error} presents the empirical generalization gap of $p$-LPQKs with $p = 10, 20, 50, 100, 150, 200, 252$ and the GFQK in the same classification setting as Fig.~\ref{Fig:Kernel-bandwidth-results} for $C = 5.0$ and $\gamma = 0.2$. The empirical generalization gap is estimated by the difference between the testing and training accuracy and it informs us about the magnitude of the complexity term. We observe the increasing generalization gap with $p$ in our numerical example, as predicted by Theorem~\ref{Theorem:Generalization-pTIQK}. However, it is important to note the non-trivial role of the kernel bandwidth in the generalization bound. It would therefore be interesting to develop a bound that takes both $p$ and $\lambda$ into account.

\section{Discussion}
\label{Sec:Conclusion}

Fundamental understanding of QML is necessary to pave a way for a practical quantum advantage. In this work, we focused on quantum kernel methods and studied a fundamental connection between different trace-induced kernels. We proposed a unified framework for generalized trace-induced quantum kernels (GTQKs) that encompasses existing kernels in the literature including the global fidelity quantum kernel (GFQK) and linear projected quantum kernels (LPQKs). Specifically, given an arbitrary orthonormal Hermitian basis $\AC = \{ A_i \}_i$, the fundamental building blocks, Lego kernels, can be constructed and the GTQK is a positive linear combination of these Lego kernels. This allows a fair comparison of expressivity and generalizability between different classes of the GTQKs through the number of non-zero weights $p$. In addition, we analyzed the role of these weights in the regularized training of the model and revealed an the inductive bias imposed towards the base functions with large associated weights. We also considered the Mercer eigenbasis, where Lego kernels form an orthornormal RKHS. We thus demonstrated a hierarchical structure of expressivity for these quantum kernel models.

Next we studied practicalities of the unified framework when applied to kernel models constructed in the Pauli basis. We proposed a systematic approach to naturally increase the complexity of such models, leading to a new version of LPQKs, $H$-body LPQKs. Training with the $H$-body LPQKs for fixed $H$ requires less quantum resources in general compared to the GFQK. The LPQKs with fixed $H$ have a favorable measurement shot scaling of $\OC(N)$ compared to $\OC(N^2)$ for the GFQKs. Moreover, LPQKs can be implemented using shallower quantum circuits. Through an empirical study using the fashion-mnist dataset, we found that the LPQKs can achieve similar prediction accuracy to the GFQK, which has larger expressivity. Our study thus provides theoretical and empirical evidence in favor of using LQPKs over the GFQK.

Our work contributes to understanding how the form of a kernel function itself can affect different aspects of QML models including expressivity, inductive bias, generalizability and resource requirements. There remains many open question worth further investigation in this direction. Is it possible to find a fundamental connection between GTQKs and other forms of quantum kernels? How can symmetries of the data be incorporated into this approach to build the optimal kernel function? Finally, generalization of the present approach beyond the special case of supervised learning to other learning problems is another important direction for future work.

\begin{acknowledgments}

This research is supported by the National Research Foundation, Singapore and A*STAR under its CQT Bridging Grant and Quantum Engineering Programme NRF2021-QEP2-02-P02, A*STAR(\#21709). ST is later supported by the Sandoz Family Foundation-Monique de Meuron
program for Academic Promotion and partially by Thailand Science Research and Innovation Fund Chulalongkorn University (IND66230005).

\end{acknowledgments}

\bibliography{bibliography.bib}

\clearpage
\newpage
\onecolumngrid

\appendix
\vspace{0.5in}
\begin{center}
	{\Large \bf Appendix} 
\end{center}

\section{Generalization bounds for binary classification}
\label{Appendix-Sec:Generalization-bounds-introduction}

In this Appendix, we will briefly review generalization bounds for linear binary classifiers, e.g.: support vector machines as an example. The introduction in this section is based on Ref.~\cite{mohri2018foundations} so motivated readers can refer to for more rigorous proofs.

\subsection{Generalization bound via Rademacher complexity}

Let $\mathcal{X}$ be the input space and $\mathcal{Y}$ be the label set relating via the concept $c : \mathcal{X} \rightarrow \mathcal{Y}$. 
Here, we are considering binary classification problems, hence the label set only contains two values, i.e.: $\mathcal{Y} = \{0,1\}$. The aim of a learner in the supervised learning setting is to pick a hypothesis $h_S \in \HC$ in a hypothesis set $\HC$ based on the labelled sample $S=\{(\vec{x}_1 ,c(\vec{x}_1)), \dots, (\vec{x}_N, c(\vec{x}_N))\}$ drawn according to an independently and identically distributed (i.i.d) $\vec{x}_i \sim \mathcal{D}$ that has a small generalization error with respect to the target concept $c$. To quantify the generalization performance for binary classification on data $x$ sampled from the data distribution $\mathcal{D}$, we introduce the $0$-$1$ loss function,
\begin{align} \label{Eqn:0-1_lossfn}
    \LC_{h} = \mathbb{P}_{\vec{x} \sim \mathcal{D}} [h(\vec{x}) \ne c(\vec{x})] = \mathbb{E}_{\vec{x} \sim \mathcal{D}}[1_{h(\vec{x})\ne c(\vec{x})}],
\end{align}
where $h(\vec{x})$ and $c(\vec{x})$ are the hypothesis and target concepts, respectively. Operationally, this loss function counts the expected number of incorrect predictions, and the goal is to find a classifier that minimizes this expected error.

However, the generalization error of a hypothesis is not directly accessible to the learner, since both the distribution $\mathcal{D}$ and the target concept $c$ are unknown. Hence, one can only estimate the generalization error based on the labeled samples $S=\{ (\boldsymbol{x}_i,y_i)\}_{i=1}^N$, where we denote $y_i = c(\vec{x}_i)$. The estimated generalization error using $S$ is known as the empirical error or empirical risk of $h$ and is defined as
\begin{align}
     \LC_{h}(S)  = \frac{1}{N} \sum_{i=1}^N 1_{h(\vec{x}_i)\ne c(\vec{x}_i)}.
\end{align}
One of the goals of statistical learning theory is to find the generalization bound $\mathcal{B}$ of the difference between the real and empirical risk, i.e: $\LC_{h} - \LC_{h}(S) \le \mathcal{B}$. For binary classification, one can use the Rademacher complexity (Theorem 3.5 in Ref.~\cite{mohri2018foundations}) to bound the generalization error.

The empirical Rademacher complexity for sample $S$ is defined as
\begin{align}
    \mathfrak{R}_S(G) = \mathbb{E}_{\bm \sigma} \Bigg[\sup_{g\in G} \frac{1}{N} \sum_{i=1}^N \sigma_i g(\vec{x}_i,y_i) \Bigg],
\end{align}
where ${\bm \sigma} = (\sigma_i,\dots,\sigma_N)^T$ is the vector of independent uniform random variables that take values in $\{-1,+1\}$, $G = \{g: (\vec{x},y) \rightarrow \ell(h(\vec{x}),y): h \in \mathcal{H}\}$ is the family of loss functions associated with $\HC$ that maps $\mathcal{Z} = \mathcal{X} \times \mathcal{Y}$ to $\mathbb{R}$, and $\ell : \mathcal{Y} \times \mathcal{Y} \rightarrow \mathbb{R}$ is an arbitrary loss function.
The Rademacher complexity captures the richness of a family of functions by measuring the degree to which a hypothesis set can fit random noise, i.e: the more complex the family $G$ is, the better it correlate with random noise, on average. By averaging the empirical Rademacher complexity over all samples of size $N$ drawn according to $\mathcal{D}$, one can obtain the Rademacher complexity $\mathfrak{R}_N(G)$
\begin{align}
    \mathfrak{R}_N(G) = \mathbb{E}_{S\sim \mathcal{D}^N} [\mathfrak{R}_N(G)].
\end{align}
In contrast to the empirical Rademacher complexity, the Rademacher complexity does not depend on the sample $S$. To bound the generalization error for binary clasification problem, one can set the loss function $\ell$ as the 0-1 loss function and bound the generalization error $\LC_{h}$ using the associated (empirical) Rademacher complexity. 

\begin{theorem} [Rademacher complexity bounds - binary classification (Theorem 3.5 in Ref.~\cite{mohri2018foundations})] Let $\HC$ be a family of functions taking values in $\{-1,+1\}$ and let $\mathcal{D}$ be the distribution over the input space $\mathcal{X}$. Then, for any $\delta > 0$, with probability at least $1-\delta$ over a sample $S$ of size $N$ drawn according to $\mathcal{D}$, each of the following holds for any $h\in \HC$
    \begin{align} \nonumber
    \LC_{h} &\le \LC_{h}(S) + \mathfrak{R}_N(\HC) + \sqrt{\frac{\log \frac{1}{\delta}}{2N}} \\ \label{Apendix-eqn:General-GB}
    \mathrm{and} \quad 
    \LC_{h} &\le \LC_{h}(S) + \hat{\mathfrak{R}}_S (\HC) + 3\sqrt{\frac{\log \frac{2}{\delta}}{2N}}.
\end{align} 
\end{theorem}

\subsection{Linear classifiers and their generalization bounds}
\label{Appendix-Subsec:Linear-Classifiers-GB}
A possible hypothesis set for binary classification is the linear classifier, which is defined as 
\begin{align}\label{Appendix-eqn:Hypothesis-class-linear}
    \HC = \{ \boldsymbol{x} \rightarrow \text{sign}({\bf w} \cdot \boldsymbol{x} + b): {\bf w} \in \mathbb{R}^N, b \in \mathbb{R} \},
\end{align}
where ${\bf w}$ is the normal vector to the hyperplane given by ${\bf w} \cdot \boldsymbol{x} + b = 0$ with scalar $b$. The hypothesis set $\HC$ labels the points positively (negatively) if it located on one side (the other side) of the  hyperplane. For a linearly separated training sample $S$, there exist $({\bf w}, b) \in (\mathbb{R}^N - \{{\bf 0}\} \times \mathbb{R})$ such that $\forall i \in [N], y_i({\bf w} \cdot \boldsymbol{x}_i + b) \ge 0$, and the support vector machine solution is the separating hyperplane with the maximum geometric margin, i.e: $\left \{ \CC_h = \min_{i \in [N]} \frac{|{\bf w} \cdot \boldsymbol{x}_i + b|}{||{\bf w}||_2} \right\}$, i.e.: the maximum-margin hyperplane. One can then find the optimal ${\bf w}$ and $b$ by solving the primal problem
\begin{align}
    \min_{{\bf w}, b} & ~\frac{1}{2} ||{\bf w}||^2\\
    \text{subject to:} & ~y_i ({\bf w \cdot \vec{x}_i + b}) \ge 1, \forall i \in [N]
\end{align}
or the dual optimization problem
\begin{align}
    \max_{{\bm \alpha}} &~ \sum_{i=1}^N \alpha_i - \frac{1}{2} \sum_{i,j=1}^N \alpha_i \alpha_j y_i y_j (\boldsymbol{x}_i \cdot \boldsymbol{x}_j)\\ 
    \text{subject to:} & ~\alpha_i \ge 0 \wedge \sum_{i=}^N \alpha_i y_i = 0, \forall i \in [N]. 
\end{align}
where $\alpha_i \ge 0$. For the case of a non-separable dataset, one introduces slack variables $\xi_i$ that measure the distance by which vector $\boldsymbol{x}_i$ violates the desired inequality, $y_i ({\bf w}\cdot \boldsymbol{x}_i +b) \ge 1$, hence, turning the primal optimization into
\begin{align}
    \min_{{\bf w}, b, {\bm \xi}} & ~\frac{1}{2} ||{\bf w}||^2 + C \sum_{i=1}^N \xi_i^p\\
    \text{subject to:} & ~y_i ({\bf w} \cdot \boldsymbol{x}_i + b) \ge 1 - \xi_i \wedge \xi_i \le 0, \forall i \in [N],
\end{align}
where the first term aims to maximize the margin while the second term tries to minimize the total amount of slack. The corresponding dual problem becomes
\begin{align} \nonumber
    \max_{{\bm \alpha}} &~ \sum_{i=1}^N \alpha_i - \frac{1}{2} \sum_{i,j=1}^N \alpha_i \alpha_j y_i y_j (\boldsymbol{x}_i \cdot \boldsymbol{x}_j)\\ \label{Appendix-eqn:Soft-linear-SVM}
    \text{subject to:} & ~0 \le \alpha_i \le C \wedge \sum_{i=}^N \alpha_i y_i = 0, \forall i \in [N]. 
\end{align}
where, compared to the separable dataset set case, the $\alpha_i$ is upper bound by $C$. Both dual optimization problems give rise to the same solution $h(\boldsymbol{x}) = \text{sgn}({\bf w} \cdot \boldsymbol{x} + b) = \text{sgn} \left (\sum_{i=1}^N \alpha_i y_i (\boldsymbol{x}_i \cdot \boldsymbol{x}) + b \right)$ with $b = y_i - \sum_{j=1}^N \alpha_j y_j (\boldsymbol{x}_j \cdot \boldsymbol{x}_i)$ and the only difference between the two solutions is the upper bound of $\alpha$.

\subsubsection{Generalization bound via confidence margin}

Alternatively, one can formulate the linear binary classification problem based on the confidence margin $yh(\boldsymbol{x})$ for any $h \in \HC$, where $h$ classifies $\boldsymbol{x}$ correctly with confidence $|h(\boldsymbol{x})|$ when $y h(\boldsymbol{x}) > 0$, instead of the geometric margin that gives rise to the support vector machine. The geometric and the confidence margins are related in the separable case by $|yh(\boldsymbol{x})| \ge \CC_{geom} ||{\bf w}||$. The confidence margin of a real-valued function $h$ at a point $\vec{x}$ labeled with $y$ is the quantity $y h(\vec{x})$ and the associated margin loss function is defined as 
\begin{align}
    \Phi_\CC(z) = \begin{cases}
        1 & \text{if} \quad z \le 0\\
        1- \frac{z}{\CC} & \text{if} \quad 0 \le z \le \CC\\
        0 & \text{if} \quad \CC \le z
    \end{cases},
\end{align}
with the empirical margin loss defined as 
\begin{align}
     \LC_{h}^{(\CC)}(S) = \frac{1}{N} \sum_{i=1}^N \Phi_\CC(y_i h(\vec{x}_i)) \le \frac{1}{N} \sum_{i=1}^N 1_{y_i h(\vec{x}_i) \le \CC}
\end{align}
where the upper bound follows from $\Phi_\CC(y_i h(\vec{x}_i)) \le 1_{y_i h(\vec{x}_i) \le \CC}$ for any $i \in [m]$. Given this empirical margin loss, one can then find the (empirical) Rademacher complexity for the hypothesis class $\HC$ in Eq.~\eqref{Appendix-eqn:Hypothesis-class-linear} and use it to bound the generalization error
\begin{align} 
    \LC_{h}^{(\CC)} &\le \LC_{h}^{(\CC)}(S) + \frac{2}{\CC}\mathfrak{R}_N(\HC) + \sqrt{\frac{\log \frac{1}{\delta}}{2N}} \\ \label{Appendix-eqn:Margin-GB}
    \mathrm{and} \quad 
    \LC_{h}^{(\CC)} &\le \LC_{h}^{(\CC)}(S) + \frac{2}{\CC}\hat{\mathfrak{R}}_S (\HC) + 3\sqrt{\frac{\log \frac{2}{\delta}}{2N}}.
\end{align}
If both the weight vector ${\bf w}$ and data vector $\boldsymbol{x}$ of linear hypothesis of $\HC$ are bounded, i.e.: $||{\bf w}|| \le \Lambda$ and $||\boldsymbol{x}|| \le R$ for $\Lambda, R \ge 0$, one can further bound the empirical Rademacher complexity by $\hat{\mathfrak{R}}_S (\HC) \le \sqrt{\frac{\Lambda^2 R^2}{N}}$. The generalization bound for $\HC$ based on the empirical Radamader complexity becomes
\begin{align}
    \LC_{h}^{(\CC)} &\le \LC_{h}^{(\CC)}(S) + \frac{2}{\CC}\sqrt{\frac{\Lambda^2 R^2}{N}} + 3\sqrt{\frac{\log \frac{2}{\delta}}{2N}}.
\end{align}
Hence, to achieve small generalization bound, one has to make sure that both the empirical margin loss and $\CC/(\Lambda R)$ are small. This happens when $\CC$ is relatively large while few points are either classified incorrectly or correctly, but with margin less than $\CC$.

\subsubsection{Justification for margin-maximization algorithms}

As the margin parameter $\CC$ is a free parameter it must be selected beforehand, but one can make the bound in Eq.\eqref{Appendix-eqn:Margin-GB} hold uniformly for all $\CC \in (0,r]$ at the cost of an additional term $\sqrt{\frac{\log \log_2 \frac{2r}{\CC}}{N}}$. Hence, for any $\delta > 0 $, with probability at least $1-\delta$, the following holds for all $h \in \{\boldsymbol{x} \mapsto {\bf w} \cdot \boldsymbol{x}: ||{\bf w}|| \le 1\}$ and $\CC \in (0,r]$
\begin{align}
    \LC_{h}^{(\CC)} &\le \LC_{h}^{(\CC)}(S) + \frac{4}{\CC}\mathfrak{R}_N(\HC) + \sqrt{\frac{\log \log_2 \frac{2r}{\CC}}{N}} + \sqrt{\frac{\log \frac{2}{\delta}}{2N}} \\
    \mathrm{and} \quad 
    \LC_{h}^{(\CC)} &\le \LC_{h}^{(\CC)}(S) + \frac{4}{\CC}\hat{\mathfrak{R}}_S (\HC) + \sqrt{\frac{\log \log_2 \frac{2r}{\CC}}{N}} + 3\sqrt{\frac{\log \frac{4}{\delta}}{2N}}.
\end{align}
Note that we have let $\Lambda = 1$ and this bound holds for $\rho$ larger than $r$. 

This bound can be used to justify the margin-maximization algorithms that gives rise to the support vector machine solution. Since the $\CC$-margin loss function is upper bounded by the $\CC$-hinge loss
\begin{align}
    \forall z \in \mathbb{R}, \Phi_\CC (z) = \min \Big(1,\max \Big(0,1-\frac{z}{\CC} \Big) \Big) \le \max \Big(0,1-\frac{z}{\CC} \Big),
\end{align}
and $h/\CC$ has the same generalization error as $h$, one can derive a generalization bound for $h \in \{\boldsymbol{x} \rightarrow {\bf w} \cdot \boldsymbol{x}: ||{\bf w}|| \le 1/\CC\}$
\begin{align}
    \LC_{h}^{(\CC)} \le \frac{1}{N} \sum_{i=1}^N \max(0,1-y_i ({\bf w} \cdot \boldsymbol{x}_i)) + \frac{4}{\CC}\sqrt{\frac{R^2}{N}} + \sqrt{\frac{\log \log_2 \frac{2r}{\CC}}{N}} + \sqrt{\frac{\log \frac{2}{\delta}}{2N}} 
\end{align}
for all $\CC >0$, using it to construct an algorithm that selects $\{\bf w \}$ and $\CC > 0$ to minimize the right-hand side. By letting $\CC$ be a free parameter of the algorithm and optimizing only $\{\bf w \}$, one could keep only the first term on the right hand side of the bound and construct the following optimization algorithm that coincides with support vector machine to select ${\bf w}$
\begin{align}
    \min_{||{\bf w}||^2 \le \frac{1}{\CC^2}} \frac{1}{N} \sum_{i=1}^N \max(0,1-y_i ({\bf w} \cdot \boldsymbol{x}_i)) \quad \rightarrow \quad \min_{{\bf w}} \lambda ||{\bf w}||^2 + \frac{1}{N} \sum_{i=1}^N \max(0,1-y_i ({\bf w} \cdot \boldsymbol{x}_i))
\end{align}
where the second minimization problem is obtained by turning the first optimization problem into a Lagragian problem with a Lagrange variable $\lambda \ge 0$ for the constraint $||{\bf w}||^2 \le \frac{1}{\CC^2}$. For any choice of $\CC$ in the first optimization problem, there exists an equivalent dual variable $\lambda$ in the second problem that achieves the same optimal ${\bf w}$.

\subsection{Nonlinear classifiers and their generalization bounds}

The linear classifiers introduced in Appendix~\ref{Appendix-Subsec:Linear-Classifiers-GB} will perform poorly if the dataset is nonlinear. By exploiting the fact that the hypothesis solution in Eq.~\eqref{Appendix-eqn:Soft-linear-SVM} only depends on inner products between vectors and not directly on the vectors themselves, one can replace $\boldsymbol{x}_i$ by a feature map $\Phi(\boldsymbol{x}_i)$, turning the dual optimization into
\begin{align}
    \max_{{\bm \alpha}} &~ \sum_{i=1}^N \alpha_i - \frac{1}{2} \sum_{i,j=1}^N \alpha_i \alpha_j y_i y_j k(\boldsymbol{x}_i, \boldsymbol{x}_j)\\
    \text{subject to:} & ~0 \le \alpha_i \le C \wedge \sum_{i=}^N \alpha_i y_i = 0, \forall i \in [N]. 
\end{align}
where $k(\boldsymbol{x}_i, \boldsymbol{x}_j) = \langle \Phi(\boldsymbol{x}_i) , \Phi(\boldsymbol{x}_j) \rangle$ is the inner product between feature maps $\Phi(\boldsymbol{x}_i)$ and $\Phi(\boldsymbol{x}_j)$ for data points $\boldsymbol{x}_i$ and $\boldsymbol{x}_j$, respectively. The hypothesis solution can then be written as $h(\vec{x}) = \text{sgn} \Big( \sum_{i=1}^N \alpha_i y_i k(\vec{x}_i,\vec{x}) + b \Big)$.

Now we will utilize the Rademacher complexity-based generalization bound to bound the generalization error of the hypothesis based on feature map $\Phi$, i.e: $\HC = \{\vec{x} \mapsto \langle {\bf w}, {\bm \Phi}(\vec{x}) \rangle : ||{\bf w}||_{\mathcal{H}} \le \Lambda\}$, where $\mathcal{H}$ is the feature map associated with $k$. One can use the classical kernel theory developed in Appendix~\ref{Appendix-Sec:Intro-RKHS} to study the properties of the kernel and its associated model class. For $S \subseteq \{\vec{x}: k(\vec{x},\vec{x}) \le R^2 \}$, the empirical complexity is bounded by 
\begin{align} \label{eq:empirical-rademacher-kernel}
    \hat{\mathfrak{R}}_S (H) \le \frac{\Lambda \sqrt{\tr[K]}}{N} \le \sqrt{\frac{\Lambda^2 R^2}{N}},
\end{align}
where $K$ is the kernel matrix with matrix elements $K_{ij} = k(\vec{x}_i,\vec{x}_j)$. For fixed $\CC > 0$ and $\delta >0$, the following holds with probability at least $1-\delta$ for any $h \in \{\boldsymbol{x} \mapsto {\bf w} \cdot {\bm \Phi}(\boldsymbol{x}): ||{\bf w}||_{\mathcal{H}} \le \Lambda \}$
\begin{align}
    \LC_{h}^{(\CC)} &\le \LC_{h}^{(\CC)}(S) + \frac{2}{\CC}\sqrt{\frac{\Lambda^2 R^2}{N}} + \sqrt{\frac{\log \frac{1}{\delta}}{2N}} \\
    \mathrm{and} \quad 
    \LC_{h}^{(\CC)} &\le \LC_{h}^{(\CC)}(S) + \frac{2}{\CC}\frac{\sqrt{\Lambda^2 \tr[K]}}{N} + 3\sqrt{\frac{\log \frac{2}{\delta}}{2N}}.
\end{align}
Hence, the trace of the kernel matrix can be used to tune the complexity of hypothesis set and hence control the generalization bound.

\section{An introduction to reproducing kernel Hilbert space}
\label{Appendix-Sec:Intro-RKHS}

Reproducing Kernel Hilbert Space (RKHS) is one of the core elements in the classical kernel theory. It enables us to analyze the learning ability of kernel-based machine learning models. In this Appendix, we review the necessary background of the theory of RKHS required to understand our analysis on the class of generalized trace-induced quantum kernels. For brevity, we will omit the proofs of standard kernel theory results here. Interested reader can refer to standard textbooks for more information~\cite{scholkopf2002learning,steinwart2008support}. First, we will give the definitions on various important quantities used in kernel theory.

\begin{definition} [Kernel matrix] Given a kernel function $k : \mathcal{X} \times \mathcal{X} \rightarrow \mathbb{R}$ and pattern $x_1, \dots, x_N \in \XC$, the $N \times N$ matrix $K$ with elements $K_{ij} = k(x_i,x_j)$ is called the kernel matrix (or Gram matrix) of $k$ with respect to $x_1, \dots, x_N$.
\end{definition}

\begin{definition} [Positive semi-definite matrix] A real symmetric $N \times N$ matrix $K$ satisfying 
\begin{align}
    \sum_{ij} c_i c_j K_{ij} \ge 0
\end{align}
for all $c_i \in \mathbb{R}$ is called positive definite. A symmetric matrix is positive definite if and only if its eigenvalues are non-negative.
    
\end{definition}

\begin{definition} [Positive semi-definite (Mercer) kernel] Let $\XC$ be a non-empty set. A function $k$ on $\XC \times \XC$ which for all $N \in \mathbb{N}$ and all $x_1,\dots,x_N \in \XC$ give rise to a positive definite Gram matrix is called a positive definite kernel.
\end{definition}

\subsection{Constructing reproducing kernel Hilbert space via Moore-Aaronjain construction}
\label{Appendix-Subsec:RKHS_MA-construction}

There are various ways to construct a reproducing kernel Hilbert space. One standard procedure is to construct a feature space directly from a kernel, i.e. the Moore-Aaronjain construction.

\textbf{(1) Define a map $\Phi(\cdot,\vec{x}) = k(\cdot,\vec{x})$ by fixing one domain of the kernel $k(\vec{x'},\vec{x})$.} Given a real-valued positive semi-definite kernel $k$ and nonempty set $\mathcal{X}$, one can define the map from $\mathcal{X}$ into a space of functions that map $\mathcal{X}$ into $\mathbb{R}$
\begin{align}
    \Phi: & \mathcal{X} \rightarrow \mathbb{R}\\
    & \vec{x} \rightarrow k(\cdot, \vec{x}).
\end{align}
The map $\Phi$ assigns the value $k(\vec{x},\vec{x}')$ to $\vec{x}'\in \mathcal{X}$, turning each pattern into a function in domain $\mathcal{X}$. In other words, the pattern is now represented by its similarity to all other point on the input domain.  

\textbf{(2) Construct a inner product space by using the image of $\Phi$.} Given the images of $\Phi$ $\{ k(\cdot,\vec{x}) \: |\: \vec{x}\in \mathcal{X}\}$, i.e.: the spanning set, one can construct a vector space by linear combining functions in this set. This forms a vector space of functions
\begin{align}
   \mathcal{V} := \left \{\sum_{i=1}^N \alpha_i k(\cdot, \vec{x}_i) \: | \: \alpha_i \in \mathbb{R}, \: N\in \mathbb{N}, \: \vec{x}_i \in \mathcal{X} \right \}.
\end{align}
The vector space $\mathcal{V}$ can be promoted to an inner product space $\mathcal{G}$ by endowing it with an inner product
\begin{align}
    \langle f, g\rangle_{\mathcal{G}} := \sum_{i=1}^N \sum_{j=1}^{N'} \alpha_i \beta_j k(\vec{x}_i,\vec{x}_j),
\end{align}
where $f(\cdot)=\sum_{i=1}^N \alpha_i k(\vec{x}_i,\cdot)$ and $g(\cdot) =  \sum_{j=1}^{N'} \beta_j k(\vec{x}'_j,\cdot)$, with $N,N'\in \mathbb{N}$, $\alpha_i, \beta_j \in \mathbb{R}$, and $\vec{x}_i, \vec{x}'_j \in \mathcal{X}$. This inner product is well-defined as $\langle f, g\rangle_{\mathcal{G}} = \sum_j \beta_j f(\vec{x}'_j) = \sum_i \alpha_i g(\vec{x}_i)$, i.e.: it is independent of the representation of both $g$ and $f$. One can recover $f(\vec{x})$ by setting $g(\cdot) = k(\cdot,\vec{x})$
\begin{align}
    \langle f(\cdot), k(\cdot,\vec{x})\rangle_{\mathcal{G}} := \sum_{i=1}^N \alpha_i k(\vec{x}_i,\vec{x}) = f(\vec{x}).
\end{align}
This is known as the reproducing property of kernel $k$. 

It is easy to check that this definition satisfies the first two properties of inner product (1) bilinear and (2) symmetry. The last property of the inner product (3) $\langle f,f \rangle = 0 \rightarrow f=0$ can be shown by utilizing the reproducing property of the kernel and the Cauchy-Schwarz inequality
\begin{align}
    |f(\vec{x})|^2 = |\langle k(\vec{x},\cdot), f \rangle|^2 \le k(\vec{x},\vec{x}) \langle f,f \rangle,
\end{align}
where the inequality is obtained by invoking the Cauchy-Schwarz inequality. Hence, $\langle f,f \rangle = 0 $ implies $f=0$.

\textbf{(4) Completion of the inner product space.} To promote $\mathcal{G}$ to a proper Hilbert space, we need to take its topological completion $\bar{G}$, i.e.: we add all limits of Cauchy sequences. 

Completing step (1) - (4) is then yield a reproducing kernel Hilbert space $\mathcal{H}$ with reproducing kernel $k$. 

\begin{align}
    \mathcal{F}_{k} = \overline{\left \{ f: f(\cdot) = \sum_{i} \alpha_i k(\cdot,\boldsymbol{x}_i) \: s.t. \: ||f||_{\mathcal{F}_{k}} < \infty \right \}}.
\end{align}

\begin{definition}[Reproducing kernel] Let $\mathcal{H}$ be a Hilbert space of functions $f: \mathcal{X} \rightarrow \mathbb{R}$ defined on a non-empty set $\mathcal{X}$. A function $k: \mathcal{X} \times \mathcal{X} \rightarrow \mathbb{R}$ is called a reproducing kernel of $\mathcal{H}$ if it satisfies
\begin{enumerate}
    \item $\forall \vec{x} \in \mathcal{X}, k_{\vec{x}} = k(\vec{x},\cdot) \in \mathcal{H}$
    \item $\forall \vec{x} \in \mathcal{X}, \forall f \in \mathcal{H}, \langle f, k_{\vec{x}} \rangle_\mathcal{H} = f(\vec{x})$ (the reproducing property)
\end{enumerate}
In particular, for any $\vec{x}, \vec{x'} \in \mathcal{X}, \: k(\vec{x},\vec{x'}) = \langle k_{\vec{x}}, k_{\vec{x'}}\rangle_\mathcal{H} = \langle k_{\vec{x'}}, k_{\vec{x}} \rangle_\mathcal{H} = k(\vec{x'},\vec{x})$.
\end{definition}

\begin{definition} [Reproducing kernel Hilbert space] A Reproducing Kernel Hilbert Space (RKHS) is a Hilbert space $\mathcal{H}$ of functions $f: \mathcal{X} \rightarrow \mathbb{R}$ with a reproducing kernel $k: \mathcal{X} \times \mathcal{X} \rightarrow \mathbb{R}$ where $k(\cdot, \boldsymbol{x}) \in \mathcal{H} $ and $f(\boldsymbol{x}) = \langle f(\cdot), k(\cdot,\boldsymbol{x}) \rangle_\HC.$ The norm in RKHS is calculated as 
\begin{align}
    ||f||_\mathcal{H} := \sqrt{\langle f,f \rangle_\mathcal{H}}.
\end{align}
\end{definition}
Given a RKHS constructed in this manner, the representer theorem can be straightforwardly applied. 

\begin{definition} [Representer theorem] Let $k: \mathcal{X} \times \mathcal{X} \rightarrow \mathbb{R}$ be a positive semi-definite kernel and $\mathcal{H}$ as its corresponding RKHS. Then, for any non-decreasing function $G: \mathbb{R} \rightarrow \mathbb{R}$ and any loss function $\ell: \mathbb{R}^N \rightarrow \mathbb{R} \cup \{+ \infty \}$, the optimization problem based on training data $S = \{\boldsymbol{x}_i, y_i\}_{i=1}^N$
\begin{align}
    \underset{f \in \mathcal{H}}{\mathrm{argmin}} ~F(h) = \underset{f \in \mathcal{H}}{\mathrm{argmin}} ~G(||f||_{\mathcal{H}}) + \ell(f(\boldsymbol{x}_1), \dots, f(\boldsymbol{x}_m))
\end{align}
admits a solution of the form $f^*(\cdot) = \sum_{i=1}^N \alpha_i k(\boldsymbol{x}_i, \cdot)$. If $G$ is further assumed to be increasing, then any solution has this form. 
\end{definition}

\subsection{Mercer representation of reproducing kernel Hilbert space} \label{Appendix:Mercer-RKHS}
One of the important results from the classical kernel theory is the eigen-decomposition of the kernel. By Mercer theorem, a kernel will shares its eigenvalues and eigenfunctions with the integral operator 
\begin{align}
    (T_k f)(\boldsymbol{x}) = \int k(\boldsymbol{x},\vec{x'}) f(\vec{x'}) \mu(d\vec{x'}),
\end{align}
where $k(\vec{x}, \vec{x}') = \langle \Phi(\vec{x}), \Phi(\vec{x}') \rangle$ is the kernel constructed using the feature map $\vec{\Phi}(\cdot)$ with an associated linear model $f(\boldsymbol{x}) = {\bf w}\cdot \vec{\Phi}(\vec{x})$. Let $\{\gamma_i , \phi_i(\cdot)\}$ be the set of eigenvalues and eigenfunctions of the integral operator, i.e.: $(T_k \phi_i)(\boldsymbol{x}) = \gamma_i \phi_i(\boldsymbol{x})$. Then the kernel $k(\boldsymbol{x},\vec{x}')$ can be eigen-decomposed into
\begin{align}
    k(\boldsymbol{x},\vec{x}') = \sum_{i=1}^\infty \gamma_i \phi_i(\boldsymbol{x}) \phi_i(\vec{x}').
\end{align}
This lead to the Mercer feature map $\vec{\Psi}(\vec{x}) = (\psi_i(\vec{x}))_{i=1}^\infty$.

The linear model $f(\boldsymbol{x})$ could be then be written in term of the  Mercer feature map
\begin{align}
    f(\boldsymbol{x}) = \sum_{j=1}^\infty \tilde{\alpha}_i \psi_i(\boldsymbol{x}),
\end{align}
living in the corresponding RKHS
\begin{align}
    \mathcal{H}_{k} = \Bigg \{ f: f(\cdot) = \sum_{i=1}^{\infty} \tilde{\alpha}_i \psi_{i}(\cdot) \Bigg \},
\end{align}
with $\tilde{\alpha} \in \mathbb{R}$. Given two arbitrary functions $f(\cdot) = \sum_i \tilde{\alpha}_i \psi_{i}(\cdot)$ and $g(\cdot) = \sum_j \tilde{\beta}_j \psi_{i}(\cdot)$, the inner product in this space is defined as
\begin{align}
    \langle f, g \rangle_{\mathcal{H}_{k}}  &:= \sum_{i,j = 1}^{\infty} \tilde{\alpha}_i \tilde{\beta}_j \langle \psi_{i}(\cdot), \psi_{j}(\cdot) \rangle_{\mathcal{H}_{k}} = \sum_{i=1}^{\infty} \tilde{\alpha}_i \tilde{\beta}_i,
\end{align}
in which the kernel still has the reproducing property $f(\boldsymbol{x}) = \langle f, k(\cdot, \boldsymbol{x}) \rangle_{\mathcal{H}_{k}}$, enforcing the orthogonality condition of $\psi_i(\cdot)$, i.e.: $\langle \psi_{i}(\cdot), \psi_{j}(\cdot) \rangle_{\mathcal{H}_{k}} = \delta_{ij}$.

Instead of the the Mercer feature map, one could construct the RKHS using eigenfunctions $\phi_{i}(\cdot)$
\begin{align}
    \mathcal{H}'_{k} = \Bigg \{ f': f'(\cdot) = \sum_{i=1}^{\infty} \tilde{\alpha}_i \phi_{i}(\cdot) \Bigg \},
\end{align}
where the inner product between two arbitrary functions in $\mathcal{H}'_{k}$, $f' = \sum_i \tilde{\alpha}_i \phi_{i}(\cdot)$ and $g' = \sum_i \tilde{\beta}_i \phi_{i}(\cdot)$ is defined as
\begin{align}
    \langle f', g' \rangle_{\mathcal{H}'_{k}}  := \sum_{i=1}^{\infty} \frac{\tilde{\alpha}_i \tilde{\beta}_i}{\gamma_{i}},
\end{align}
with $\tilde{\alpha}_i, \tilde{\beta}_i \in \mathbb{R}$. Similarly, the kernel has a reproducing property $f'(\boldsymbol{x}) = \langle f'(\cdot), k(\cdot, \boldsymbol{x}) \rangle_{\mathcal{H}'_{k}}$. 

As the RKHS is uniquely determined by the kernel and vice versa, hence, difference feature spaces constructed from the same kernels are isometric isomorphic to each other. Invoking the representer theorem yield the same linear model as Sec.~\ref{Appendix-Subsec:RKHS_MA-construction}. Notice that the integral operator shares its eigenfunctions and eigenvalues with the associated kernel, hence the effect of the integral operator on an arbitrary function $f'(\cdot) = \sum_i \tilde{\alpha}_i \phi_{i}(\cdot)$ is just a re-scaling of $\tilde{\alpha}_i$ by the corresponding eigenvalues,
\begin{align}
    (T_{k}f')(\boldsymbol{x}) = \sum_{i=1}^\infty \tilde{\alpha}_i \gamma_{i}\phi_{i}(\boldsymbol{x}).
\end{align}
The operator $T_{k}^{1/2}$ for which $T_{k} = T_{k}^{1/2} \circ T_{k}^{1/2}$ then re-scales the parameter $\tilde{\alpha}_i$ by $\sqrt{\gamma_{i}}$. Hence, $T_{k}^{1/2}$ induces an isometric isomorphism between $\mathcal{H}_{k}$ and $\mathcal{H}'_{k}$. 

\subsection{Operations with kernels}
\label{Appendix-Subsec:Kernel-operations}
We know how to construct the kernels and their corresponding reproducing kernel Hilbert space. Now, we will provide a few ingredients to construct new kernels based on existing kernels. Without loss of generality, we consider the construction of new kernels from two kernels, but the generalization to multiple kernels is straightforward. We omitted the proofs, but they can be found in standard textbooks.

\begin{lemma} [Sum and scaling of kernels\label{Lemma:Sum-scaling-kernels}] If $k,\: k_1$, and $k_2$ are kernels on $\mathcal{X}$, and $\alpha \ge 0$ is a scalar, then $\alpha k$, $k_1 + k_2$ are kernels.
\end{lemma}

The kernel $k = k_1 + k_2$ is a valid kernel by Lemma.~\ref{Lemma:Sum-scaling-kernels}, hence it is associated with a RKHS $\mathcal{H}_k$, and $\mathcal{H}_k$ can be constructed from $\mathcal{H}_{k_1}$ and $\mathcal{H}_{k_2}$.

\begin{theorem}[Sum of RKHSs] Let $k_1, k_2 \in \mathbb{R}_+$ and $k = k_1 + k_2$. Then
\begin{align}
    \mathcal{H}_k = \mathcal{H}_{k_1} + \mathcal{H}_{k_2} = \{f_1 + f_2: f_1 \in \mathcal{H}_{k_1}, f_2 \in \mathcal{H}_{k_1}\} 
\end{align}
and $\forall f \in \mathcal{H}_{k}$
\begin{align}
    ||f||_{\mathcal{H}_{k}}^2 = \min_{f = f_1 + f_2} \Big \{ ||f||_{\mathcal{H}_{k_1}}^2 + ||f||_{\mathcal{H}_{k_2}}^2\Big\}.
\end{align}
If $\mathcal{H}_{k_1} \cap \mathcal{H}_{k_2} = 0$ the norm reduces to
\begin{align}
    ||f||_{\mathcal{H}_{k}}^2 = ||f||_{\mathcal{H}_{k_1}}^2 + ||f||_{\mathcal{H}_{k_2}}^2.
\end{align}
\end{theorem}

Similarly, a product of kernels is still a valid kernel.

\begin{theorem}[Products of kernels] Let $k_1$ and $k_2$ be kernels on $\mathcal{X}$ and $\mathcal{Z}$, respectively. Then
\begin{align}
    k((\vec{x},\vec{z}),(\vec{x'},\vec{z'})) := k_1(\vec{x},\vec{x'})k_2(\vec{z},\vec{z'})
\end{align}
is a kernel on $\mathcal{X} \times \mathcal{Z}$. In addition, there is an isometric isomorphism between $\mathcal{H}_k$ and the Hilbert space tensor product $\mathcal{H}_{k_\mathcal{X}} \otimes \mathcal{H}_{k_\mathcal{Z}}$. In addition, if $\mathcal{X} = \mathcal{Z}$,
\begin{align}
    k(\vec{x},\vec{x'}) := k_1(\vec{x},\vec{x'})k_2(\vec{x},\vec{x'})
\end{align}
is a kernel on $\mathcal{X}$.
\end{theorem}

\subsection{Multiple kernel learning theory}\label{Appendix:multiple-kernel-learning}
Based on the operations introduced in Appendix.~\ref{Appendix-Subsec:Kernel-operations}, one can construct new kernels using linear combinations of a set of pre-chosen kernels. Specifically, given a fixed of set $p>1$ kernels $k_1, \dots, k_p$, one can construct a new kernel as linear combination of the kernels with non-negative weights $w_i$,
\begin{align}
    k = \sum_{i=1}^p w_i k_i,
\end{align}
with $w_i$ obeying constraint $\sum_{i=1}^q w_i^q = 1$. The non-negative constraint on the weights can be lifted if the resulting kernel is a valid kernel. For the kernel function $k$, one can find a associated feature map ${\bm \Phi}_k$ that maps the input space $\mathcal{X}$ to the reproducing kernel Hilbert space $\HC^q_p$ induced by $k$ 
\begin{align}
    \HC^q_p = \{ f(\cdot) = {\bf b} \cdot {\bm \Phi}_k (\cdot) : ||{\bf b}||_2^2 \le \Lambda\},
\end{align}
where the superscript $q$ and subscript $p$ are used to denote the weight constraint and the number of based kernels considered, respectively. For an arbitrary sample $S = \{(\vec{x}_i,y_i)\}_{i=1}^N$ of size $N$, one could use the multiple kernel learning algorithms \cite{kloft2011lp} to optimize the weights $w_i$ to find the optimal kernel for the problem of interest. In addition, the Rademacher complexity of $\HC^q_p$ can be bounded using Theorem.~\ref{Theorem:Rademacher-MKL}.

\begin{theorem} [Rademacher complexity bound for $\HC^q_p$ (Theorem 4 from Ref.~\cite{cortes2010generalization})]\label{Theorem:Rademacher-MKL}
Let $q, r \ge 1$ with $\frac{1}{q} + \frac{1}{r} = 1$ and assume that $r$ is an integer. Let $p > 1$ and assume that $k_i(\vec{x}, \vec{x}) \le R^2$
for all $\vec{x} \in \XC$ and $i \in [1,p]$. Then, for any sample $S$ of size $N$, the Rademacher complexity of the hypothesis set $\HC^2_p$ can be bounded as follows:
\begin{align} \label{Eqn:empirical-rademacher-MKL}
    \hat{\mathfrak{R}}_S (\HC^2_p) \le \frac{\sqrt{2\eta_0 ||{\bf u}||_2}}{N} \le \sqrt{\frac{2\eta_0 \sqrt{p}R^2}{N}}
\end{align}
where ${\bf u} =  (\tr[K_1],\dots,\tr[K_p])^T$ and $\eta_0 = \frac{23}{22}$. The second equality is obtained by assuming $k(\vec{x},\vec{x}) \le R^2$ for all $\vec{x} \in \mathcal{X}$.
\end{theorem}
Note that the bound in Eq.~\eqref{Eqn:empirical-rademacher-MKL} will reduce to the bound in Eq.~\ref{eq:empirical-rademacher-kernel} when considering only one base kernel and 1-norm. Interestingly, the empirical Rademacher complexity for other norms can be obtained using the method described in Ref.\cite{kloft2011lp}. Combining Eq.~\eqref{Appendix-eqn:Margin-GB} and Eq.~\eqref{Eqn:empirical-rademacher-MKL} will yield the generalization bound for the multiple kernel learning hypothesis class: For any $\delta > 0 $, with probability at least $1-\delta$, the following holds for all $h \in \HC^2_p$
\begin{align}
    \LC_{h}^{(\CC)} &\le \LC_{h}^{(\CC)}(S) + \frac{2}{\CC}\sqrt{\frac{2\eta_0 \sqrt{p}R^2}{N}} + 3\sqrt{\frac{\log \frac{2}{\delta}}{2N}}.
\end{align}

\section{Recovering existing kernels from generalized trace-induced quantum kernels}
\label{Appendix-Sec:Recover-TIQK}

In this Appendix, we will show how the existing trace-induced quantum kernels such as the global fidelity quantum kernel (GFQK) and linear projected quantum kernels (LPQKs) can be obtained from the generalized trace-induced quantum kernels by choosing an appropriate set of weights $w_i$.

\subsection{Recovering the global fidelity quantum kernel}
Setting $w_i = \frac{1}{2^n}$ in Eq.~\eqref{Eqn:GTQK-expanded} for all $i$ allows us to recover the GFQK
\begin{align}
    k_n(\boldsymbol{x}, \vec{x'}) &= \sum_{i=1}^{4^n} \tr(\rho(\boldsymbol{x})A_i)\tr(\rho(\vec{x'})A_i) = \tr(\rho(\boldsymbol{x})\rho(\vec{x'})),
\end{align}
where we utilized the decomposition of density matrix in basis $\mathcal{A}$, i.e: $\rho = \sum_{i=1}^{4^n} \tr(\rho A_i)A_i $ to obtain the second equality. The observation of GFQKs as an inner product of the feature map ${\bm \Phi}(\cdot) = (\tr(\rho(\boldsymbol{x})A_i))_{i=1}^{4^n}$, i.e: $k_n(\boldsymbol{x},\vec{x'}) =  \langle \Phi(\vec{x}), \Phi(\vec{x}') \rangle$ enables construction of different quantum feature maps for the GFQK using different decompositions. If one considers the computational basis, the feature map will be the standard feature map $\rho(\vec{x})$. For the Pauli decomposition, the feature map corresponds to the normalized Bloch vector \cite{heyraud2022noisy}, i.e: ${\bm \Phi}'(\cdot) = (\tr(\rho(\boldsymbol{x})\bar{P}^i_{\bf n}))_{i=1}^{4^n}$. While each feature map is related to the others via basis transformations, the Mercer basis $\mathcal{A}_{U_\XC}$ is preferred over the computational or Pauli basis as the functions associated with $\mathcal{A}_{U_\XC}$, i.e.: $\tr(\rho(\boldsymbol{x})A_i^{(U_\XC)})$, diagonalize the GTQK.

\subsection{Recovering linear projected quantum kernels}
To show that the GTQKs capture LPQKs, we have to first express the GTQKs in Pauli basis
\begin{align}
    k({\bm x},{\bm x'}) &= \sum_{i=1}^{4^n}w_i \tr(\rho(\boldsymbol{x})P_i)\tr(\rho({\vec{x'}})P_i)
\end{align}
where $P_i \in \{\mathbb{1},X,Y,Z\}^{\otimes n}_{\bf n} = \mathcal{P}_{\bf n}$. Note that the factor of $2^n$ in Eq.~\eqref{Eqn:GTQK-expanded} cancels with the normalization factor of the Pauli basis.

\subsubsection{s-linear projected quantum kernels}
Recall that the ${\bf s}$-linear projected quantum kernel (LPQK) is defined as $k_{\bf s}(\boldsymbol{x},\vec{x'}) =  \tr_{{\bf s}}(\rho_{\bf s}(\boldsymbol{x})\rho_{\bf s}(\vec{x'}))$ and one can always decompose the reduced density matrix $\rho_{\bf s}(\boldsymbol{x})$ in the Pauli basis, i.e.: $\rho_{\bf s} = \sum_{i=1}^{4^S} \frac{1}{2^S} \tr(\rho_{\bf s} P^{i}_{\bf s})P^{i}_{\bf s}$, where $P^{i}_{\bf s}$ are the Pauli observables for subsystem ${\bf s}$, i.e.: $P^{i}_{\bf s} \in \mathcal{P}^S_{{\bf s}} = \{\mathbb{1}, X, Y, Z\}^{\otimes S}_{{\bf s}}$ with $|\mathcal{P}^S_{{\bf s}}| = 4^S$. Note that $\mathcal{P}^S_{{\bf s}} \otimes \mathbb{1}_{\bf \bar{s}} \subset \mathcal{P}_{{\bf n}}$. Expanding ${\bf s}$-LPQKs in the Pauli basis yields
\begin{align} \label{Eqn:sLPQK-decomposition}
    k_{\bf s}(\boldsymbol{x},\vec{x'}) = \sum_{i=1}^{4^S} 
    \frac{1}{2^S} \tr(\rho_{\bf s}(\boldsymbol{x}) P^{i}_{\bf s})\tr(\rho_{\bf s}(\vec{x'}) P^{i}_{\bf s}).
\end{align}
Since $\tr_{\bf s}(\rho_{\bf s}(\boldsymbol{x}) P^{i}_{\bf s}) = \tr\left(\rho(\boldsymbol{x}) \cdot P^{i}_{\bf s} \otimes \mathbb{1}_{\bf \bar{s}}\right)$ the ${\bf s}$-LPQK can be recovered from the GTQK by setting the weights as
\begin{align}
    w_i = \begin{cases}
        \frac{1}{2^S}, & P^{i}_{\bf n}= P^{i}_{\bf s} \otimes \mathbb{1}_{\bf \bar{s}}\\
        0, & \text{otherwise}
    \end{cases}
\end{align}
Based on this framework there are a few interesting conclusions can be drawn: (1) The smallest feature map dimension that will lead to a meaningful trace-induced quantum kernels is 4, i.e.: the trace-induced quantum kernel for a single qubit system
\begin{align}
    k(\boldsymbol{x},\vec{x'}) &= \tr(\rho(\boldsymbol{x})\rho(\vec{x'})) = \frac{1}{2} \sum_{i=1}^4 \tr(\rho(\boldsymbol{x})P_i)\tr(\rho(\vec{x'})P_i),
\end{align}
where $P_i \in \{\mathbb{1}, X, Y, Z\}$. (2) The quantum models associated to $s$-LPQKs capture the quantum reservoir computing models. One might accidentally construct quantum kernel machines before if one consider a specific combination of measurement observables, e.g.: 1-RDM LPQKs. (3) The family of quantum kernels based on open quantum systems \cite{heyraud2022noisy} is a subset of ${\bf s}$-LPQKs as the quantum evolution can always be represented as the joint unitary evolution between the quantum system and the environment, followed by the tracing operation on the environment system.
\begin{align}
    \rho^{final}_{sys}(\boldsymbol{x}) = \mathcal{T}_{\boldsymbol{x}}(\rho_{sys}^{init}) = \tr_E (U(\boldsymbol{x})\rho_{sys}^{init}\otimes \rho_E U^\dagger(\boldsymbol{x})).
\end{align}

\subsubsection{S-linear projected quantum kernels}
\label{Appendix-Subsubsec:S-LPQK}
Plugging the decomposition of the ${\bf s}$-LPQK for different partitions ${\bf s}$ in Eq.(\ref{Eqn:sLPQK-decomposition}) into $k_{S}(\boldsymbol{x},\vec{x'})$ yields
\begin{align}
    k_S(\boldsymbol{x},\vec{x'}) &= \frac{1}{\sqrt{|\mathbb{S}_S|}} \sum_{{\bf s} \in \mathbb{S}_S} \sum_{i=1}^{4^S} \frac{1}{2^S} \tr_{\bf s}(\rho_{\bf s}(\boldsymbol{x}) P^{i}_{\bf s})\tr_{\bf s}(\rho_{\bf s}(\vec{x'}) P^{i}_{\bf s}).
\end{align}
This relation can be simplified by noting that there are common summands between the ${\bf s}$-LPQKs. For example, given two subsystems $\{(1,i),(1,j) ~| ~\forall i \ne j\}$, there will be terms proportional to $\tr_{1i}(\rho_{1i}(\boldsymbol{x}) \cdot X_1 \mathbb{1}_i)$ in $k_{1i}(\boldsymbol{x},\vec{x'})$ and $\tr_{1j}(\rho_{1j}(\boldsymbol{x}) \cdot X_1 \mathbb{1}_j)$ in $k_{1j}(\boldsymbol{x},\vec{x'})$ that coincide, i.e.: $\tr(\rho(\boldsymbol{x}) X_1  \mathbb{1}_{\bar{1}}) = \tr_{1i}(\rho_{1i}(\boldsymbol{x}) \cdot X_1\mathbb{1}_i) = \tr_{1j}(\rho_{1j}(\boldsymbol{x}) \cdot X_1 \mathbb{1}_j)$. Hence we can simplify $k_S(\boldsymbol{x},\vec{x'})$ by grouping common summands and breaking them down into the summation of $H$-body LPQK $k_H(\boldsymbol{x},\vec{x'})$
\begin{align}
    k_S(\boldsymbol{x},\vec{x'}) &= \frac{1}{2^S \sqrt{|\mathbb{S}_S|}} \sum_{H=0}^S \sqrt{d_H} D^{(S,H)} k_H(\boldsymbol{x},\vec{x'}), \quad 
\end{align}
where $k_H(\boldsymbol{x},\vec{x'}) = \sum_{i = 1}^{d_H} \frac{1}{\sqrt{d_H}}\tr(\rho(\boldsymbol{x}) P^i_{H}) \tr(\rho(\vec{x'}) P^i_{H})$ is the $H$-body LPQK, $D^{(S,H)} = \binom{n-H}{S-H}$ is the degeneracy of the common summands, and $P^i_H$ are $H$-body Pauli observables that act non-trivially on $H$ qubits and trivially on the remaining qubits, i.e.: $P^i_{H} \in \mathcal{P}_H = \{ \{X, Y, Z\}^{\otimes H}_{{\bf h}}\otimes \mathbb{1}_{{\bf \bar{h}}} \mid |{\bf h}| = H, \forall {\bf h}\}$. The set $\mathcal{P}_H$ contains all $H$-body Pauli observables with set size $|\mathcal{P}_H| = d_H = \binom{n}{H} \cdot 3^H$. An example of a $2$-body Pauli observable is $P_1 \otimes P_2 \otimes \mathbb{1}_{3,\dots,n}$ for $P_j \in  \{X,Y,Z\}$, while the $0$-body Pauli observables corresponds to the identity operator.

One can therefore recover the $H$-body LPQK and $S$-LPQK by setting the weights of the GTQK as 
\begin{align}
    w_i = \begin{cases}
        \frac{1}{\sqrt{d_H}}, & P^{i}_{\bf n}= P^{i}_{H}\\
        0, & \text{otherwise}
    \end{cases} \qquad \text{and} \qquad 
    w_i = \begin{cases}
        \frac{D^{(S,H)}}{2^S \sqrt{|\mathbb{S}_S|}}, & P^{i}_{\bf n}= P^{i}_{H \le S} \\
        0, & \text{otherwise}
    \end{cases},
\end{align}
respectively. The GFQK can be recovered from the $S$-LPQK by setting $S = n$. In this case, the degeneracy of the $H$-body LPQK is lifted as $D^{(n,H)} = \binom{n-H}{n-H} = 1$, where $H = \{0,1,...,n\}$, and $|\mathbb{S}_n| = 1$. Since $\bigcup_{H=0}^n \mathcal{P}^H = \mathcal{P}_{{\bf n}}$, we can write $k_n(\boldsymbol{x},\vec{x'})$ as
\begin{align}
    k_n(\boldsymbol{x},\vec{x'})
    &= \frac{1}{2^n}\sum_{H=0}^n \sqrt{d_H} k_H(\boldsymbol{x},\vec{x'})
    = \frac{1}{2^n} \sum_{i=1}^{4^n} \tr(\rho(\boldsymbol{x})P^{i}_{\bf n})\tr(\rho(\vec{x'})P^{i}_{\bf n}),
\end{align}
which is the GFQK. 

\subsubsection{H-body linear projected quantum kernels}
\label{Appendix-Subsubsec-HLPQK}
From the construction in Sec.~\ref{Appendix-Subsubsec:S-LPQK}, we identify a new type of trace-induced quantum kernel: $H$-body linear projected quantum kernels. Interestingly, the $H$-body LPQK has a corresponding quantum kernel form where it can be written in terms of the linear combinations of the weighted $S$-LPQK. Recall that the $H$-body LPQKs and $S$-LPQKs are defined as 
\begin{align}
    k_H(\boldsymbol{x},\vec{x'}) = \sum_{i = 1}^{d_H} \frac{1}{\sqrt{d_H}}\tr(\rho(\boldsymbol{x}) P^i_{H}) \tr(\rho(\vec{x'}) P^i_{H}) \quad \text{and} \quad k_S(\boldsymbol{x},\vec{x'}) &= \frac{1}{2^S \sqrt{|\mathbb{S}_S|}} \sum_{H=0}^S \sqrt{d_H} D^{(S,H)} k_H(\boldsymbol{x},\vec{x'}),
\end{align}
respectively, with $D^{(S,H)} = \binom{n-H}{S-H}$. We define the non-normalized $H$-body LPQKs and $S$-LPQKs as
\begin{align}
    k'_H(\boldsymbol{x},\vec{x'}) = \sum_{i = 1}^{d_H} \tr(\rho(\boldsymbol{x}) P^i_{H}) \tr(\rho(\vec{x'}) P^i_{H}), \quad \text{and} \quad k'_S(\boldsymbol{x},\vec{x'}) = \sum_{H=0}^S D^{(S,H)} k'_H(\boldsymbol{x},\vec{x'}),
\end{align}
respectively. We will first express $k'_H$ in terms of $k'_S$, and then use this relation to infer the normalized form. For $S=1$, we have 
\begin{align}
    k'_{S=1}(\boldsymbol{x},\vec{x'}) 
    &= \sum_{H=0}^1 D^{(1,H)} k'_H(\boldsymbol{x},\vec{x'})\\
    &=  \binom{n}{1} \cdot k'_{H=0}(\boldsymbol{x},\vec{x'}) +  \binom{n-1}{0} \cdot k'_{H=1}(\boldsymbol{x},\vec{x'})\\
    &= n k'_{H=0}(\boldsymbol{x},\vec{x'}) +  k'_{H=1}(\boldsymbol{x},\vec{x'}).
\end{align}
Note that $k'_{H=0}(\boldsymbol{x},\vec{x'})= k'_{S=0}(\boldsymbol{x},\vec{x'}) = 1$ and re-arranging the terms gives
\begin{align}\label{Appendix-Eqn:k'_H=1}
    k'_{H=1}(\boldsymbol{x},\vec{x'}) = k'_{S=1}(\boldsymbol{x},\vec{x'}) - n k'_{S=0}(\boldsymbol{x},\vec{x'}).
\end{align}
For $S=2$, we have
\begin{align}
    k'_{S=2}(\boldsymbol{x},\vec{x'}) 
    &= \sum_{H=0}^2 D^{(2,H)} k'_H(\boldsymbol{x},\vec{x'})\\
    &= D^{(2,0)} k'_{H=0}(\boldsymbol{x},\vec{x'}) + D^{(2,1)} k'_{H=1}(\boldsymbol{x},\vec{x'}) + D^{(2,2)} k'_{H=2}(\boldsymbol{x},\vec{x'})\\
    &= \frac{n(n-1)}{2} k'_{H=0}(\boldsymbol{x},\vec{x'}) + (n-1)k'_{H=1}(\boldsymbol{x},\vec{x'}) + k'_{H=2}(\boldsymbol{x},\vec{x'}).
\end{align}
Re-arranging the terms and plugging into Eq.~\eqref{Appendix-Eqn:k'_H=1} gives
\begin{align}
    k'_{H=2}(\boldsymbol{x},\vec{x'}) &= k'_{S=2}(\boldsymbol{x},\vec{x'}) - (n-1)k'_{S=1}(\boldsymbol{x},\vec{x'}) + \frac{n(n-1)}{2} k'_{S=0}(\boldsymbol{x},\vec{x'}).
\end{align}
For $S=3$, we have
\begin{align}
    k'_{S=3}(\boldsymbol{x},\vec{x'}) 
    &= \sum_{H=0}^3 D^{(3,H)} k'_H(\boldsymbol{x},\vec{x'})\\
    &= D^{(3,0)} k'_{H=0}(\boldsymbol{x},\vec{x'}) +  D^{(3,1)} k'_{H=1}(\boldsymbol{x},\vec{x'}) + D^{(3,2)} k'_{H=2}(\boldsymbol{x},\vec{x'}) +
    D^{(3,3)} k'_{H=3}(\boldsymbol{x},\vec{x'})\\
    &= \frac{n(n-1)(n-2)}{6} k'_{H=0}(\boldsymbol{x},\vec{x'}) + \frac{(n-1)(n-2)}{2} k'_{H=1}(\boldsymbol{x},\vec{x'}) + (n-2) k'_{H=2}(\boldsymbol{x},\vec{x'}) + k'_{H=3}(\boldsymbol{x},\vec{x'})\\
    &= \frac{n(n-1)(n-2)}{6} k'_{S=0}(\boldsymbol{x},\vec{x'}) + \frac{(n-1)(n-2)}{2}\cdot \big[k'_{S=1}(\boldsymbol{x},\vec{x'}) - n k'_{S=0}(\boldsymbol{x},\vec{x'})\big] \\
    & \quad + (n-2) \cdot \big[k'_{S=2}(\boldsymbol{x},\vec{x'}) - (n-1)k'_{S=1}(\boldsymbol{x},\vec{x'}) + \frac{n(n-1)}{2} k'_{S=0}(\boldsymbol{x},\vec{x'})\big] + k'_{H=3}(\boldsymbol{x},\vec{x'})\\
    &= \frac{n(n-1)(n-2)}{6} k'_{S=0}(\boldsymbol{x},\vec{x'}) - \frac{(n-1)(n-2)}{2}k'_{S=1}(\boldsymbol{x},\vec{x'}) + (n-2) k'_{S=2}(\boldsymbol{x},\vec{x'})+ k'_{H=3}(\boldsymbol{x},\vec{x'}).
\end{align}
Re-arranging the terms yields
\begin{align}
    k'_{H=3}(\boldsymbol{x},\vec{x'}) = k'_{S=3}(\boldsymbol{x},\vec{x'})  - (n-2) k'_{S=2}(\boldsymbol{x},\vec{x'}) + \frac{(n-1)(n-2)}{2}k'_{S=1}(\boldsymbol{x},\vec{x'}) - \frac{n(n-1)(n-2)}{6} k'_{S=0}(\boldsymbol{x},\vec{x'}).
\end{align}
This iterative process will yield the close form of the non-normalized $H$-body LPQK
\begin{align}
    k'_H(\boldsymbol{x},\vec{x'}) = \sum_{S=0}^{H} (-1)^{H-S} \binom{n-S}{H-S} k'_{S}(\boldsymbol{x},\vec{x'}),
\end{align}
and plugging in $k'_H = \sqrt{d_H} k_H$ and $k'_S = 2^S \sqrt{|\mathbb{S}_S|} k_S$ into this equation yields
\begin{align}
    k_H(\boldsymbol{x},\vec{x'}) = \frac{1}{\sqrt{d_H}}\sum_{S=0}^{H} (-1)^{H-S} \binom{n-S}{H-S} 2^S \sqrt{|\mathbb{S}_S|} k_{S}(\boldsymbol{x},\vec{x'}).
\end{align}
The closed form of $H$-body LPQK shows that the inner product of the quantum feature maps constructed from the $H$-body Pauli observables, i.e.: $P^i_{H} \in \mathcal{P}^H$, is just the weighted linear combination of the $S$-LPQK with $S \le H$. This observation is potentially useful when one wants to study the expressivity of quantum neural networks with measurements of all $H$-local Pauli observables.

\section{Eigendecomposition of trace-induced quantum kernels}
\subsection{Eigendecomposition of global fidelity quantum kernels}
\label{Appendix-Subsubsec:Eigendecomposition-GFQK}

In this Appendix we present the eigen-decomposition derivation in Ref.~\cite{kubler2021inductive} for completeness; an alternative derivation can be found in Ref.~\cite{canatar2022bandwidth}. The quantum model $f_n (\boldsymbol{x}) = \tr(\rho(\boldsymbol{x}) M)$ with arbitrary measurement operator $M$ is associated with the global fidelity quantum kernel $k_n(\boldsymbol{x},\vec{x'}) = \tr(\rho(\boldsymbol{x})\rho(\vec{x'}))$. The integral operator for the GFQK can be written as
\begin{align} \nonumber
    (T_{k_n}f_{n})(\boldsymbol{x}) &= \int \tr(\rho(\boldsymbol{x})\rho(\vec{x'})) \tr(\rho(\vec{x'})M) \mu(d\vec{x'})\\\nonumber
    &= \int \tr[(\rho(\vec{x'})\otimes\rho(\vec{x'}))\cdot (M \otimes \rho(\boldsymbol{x}))]\mu(d\vec{x'})\\\nonumber
    &= \tr[\int (\rho(\vec{x'})\otimes\rho(\vec{x'})) \mu(d\vec{x'})\cdot (M \otimes \rho(\boldsymbol{x}))]\\\nonumber
    &= \tr[\int (\rho(\vec{x'})\otimes\rho(\vec{x'})) \mu(d\vec{x'})\cdot (M \otimes \mathbb{1})\cdot (\mathbb{1} \otimes \rho(\boldsymbol{x}))]\\ \label{Appendix-Eqn:Integral-operator-GFQK}
    &= \tr[\tr_1[O_\mu (M \otimes \mathbb{1})] \rho(\boldsymbol{x})],
\end{align}
where $O_\mu = \int (\rho(\vec{x'})\otimes\rho(\vec{x'})) \mu(d\vec{x'})$ and $L(M) = \tr_1[O_\mu (M\otimes \mathbb{1})]$ is a linear map. As shown in Appendix C.4 of Ref.~\cite{kubler2021inductive}, $L(\cdot)$ is a linear map from Hermitian matrices to Hermitian matrices, and its eigendecomposition can be written in terms of orthonormal Hermitian operators $A_i^{(U_\XC)}$
\begin{align}
   L(M) = \sum_{i=1}^{4^n} \gamma_i \tr[A_i^{(U_\XC)}\cdot M] A_i^{(U_\XC)} ,
\end{align}
where $\gamma_i$ and $A_{i}^{(U_\XC)}$ are the eigenvalues and eigenmatrices for $L(\cdot)$, respectively. Hence, the eigenfunctions of the GFQK are $ \tr(\rho(\boldsymbol{x}) A_{i}^{(U_\XC)})$ with eigenvalue $\gamma_i$. This can be easily verified by plugging $\tr(\rho(\boldsymbol{x}) A_{i}^{(U_\XC)})$ into Eq.~\eqref{Appendix-Eqn:Integral-operator-GFQK}
\begin{align}
    \left (T_{k_n}\cdot \tr(\rho(\boldsymbol{x})A_i^{(U_\XC)}) \right )(\vec{x}) = \tr(L \left (A_i^{(U_\XC)} \right) \rho(\boldsymbol{x})) = \gamma_i \tr(\rho(\boldsymbol{x})A_i^{(U_\XC)})
\end{align}
Now, we demand that the eigenfunctions $\phi_i^{(U_\XC)}(\cdot)$ of $k_n$ be orthonormal, i.e.: $\int \phi_i^{(U_\XC)}(\boldsymbol{x}) \phi_j^{(U_\XC)}(\boldsymbol{x}) \mu(d \boldsymbol{x}) = \delta_{ij}$. One can then straightforwardly define the orthonormal eigenfunctions as $\phi_i^{(U_\XC)}(\cdot) = c \cdot \tr(\rho(\cdot) A_{i}^{(U_\XC)})$ with the normalization constant $c$ to be found via the orthogonality constraint
\begin{align}
    \int \phi_i^{(U_\XC)}(\boldsymbol{x}) \phi_j^{(U_\XC)}(\boldsymbol{x}) \mu(d \boldsymbol{x}) &= \delta_{ij}\\
    c^2 \int \tr(\rho(\boldsymbol{x}) A_{i}^{(U_\XC)}) \tr(\rho(\boldsymbol{x}) A_{j}^{(U_\XC)}) \mu(d \boldsymbol{x}) &= \delta_{ij}\\
    c^2 \tr(\int \rho(\boldsymbol{x}) \otimes \rho(\boldsymbol{x}) \mu(d \boldsymbol{x}) \cdot A_{i}^{(U_\XC)} \otimes  A_{j}^{(U_\XC)}) &= \delta_{ij}\\
    c^2 \gamma_i \tr(A_{i}^{(U_\XC)}A_{j}^{(U_\XC)}) &= \delta_{ij}.
\end{align}
Hence $c = \frac{1}{\sqrt{\gamma_i}}$ and therefore the orthonormal eigenfunctions for the GFQK are given by $\phi_i^{(U_\XC)}(\boldsymbol{x}) = \frac{\tr(\rho(\boldsymbol{x}) A_i^{(U_\XC)})}{\sqrt{\gamma_i}}$. By utilizing the orthogonality of the eigenfunctions, we can express the eigenvalues in term of $A_i^{(U_\XC)}$
\begin{align}
    \gamma_i = \int \tr(\rho(\boldsymbol{x}) A_i^{(U_\XC)})^2 \mu(d \boldsymbol{x} ).
\end{align}
Hence, $k_n$ can be written as
\begin{align}\nonumber
    k_{n}(\boldsymbol{x},\vec{x'}) &= \sum_{i=1}^{4^n}\gamma_i \phi_i^{(U_\XC)}(\boldsymbol{x}) \phi_i^{(U_\XC)}(\vec{x'})\\ \nonumber
    &= \sum_{i=1}^{4^n} \gamma_i \frac{\tr(\rho(\boldsymbol{x}) A_{i}^{(U_\XC)})}{\sqrt{\gamma_i}}  \frac{\tr(\rho(\vec{x'}) A_{i}^{(U_\XC)})}{\sqrt{\gamma_i}}\\ \label{Appendix-Eqn:Mercer-decompose-GFQK}
    &= \sum_{i=1}^{4^n} \tr(\rho(\vec{x}) A_{i}^{(U_\XC)}) \tr(\rho(\vec{x'}) A_{i}^{(U_\XC)}).
\end{align}

\subsection{Eigendecomposition of generalized trace-induced quantum kernels}
\label{Appendix-Subsubsec:Eigendecomposition-GTQK}
The generalized trace-induced quantum kernel (GTQK) in Mercer basis $A_{i}^{(U_\XC)}$ is defined as 
\begin{align}\label{Appendix-eq:GTQK-eigenform-expanded}
    k(\boldsymbol{x},\vec{x'}) = \tr(\Tilde{\rho}(\boldsymbol{x})\Tilde{\rho}(\vec{x'}))
    = \sum_{i=1}^{4^n} 2^n w_i \tr(\rho(\vec{x}) A_{i}^{(U_\XC)}) \tr(\rho(\vec{x'}) A_{i}^{(U_\XC)}),
\end{align}
where $\Tilde{\rho}(\cdot) = \sum_i \tr(\rho(\cdot)\tilde{A}_i)\tilde{A}_i$ with $\tilde{A}_i = \sqrt[\leftroot{-2}\uproot{2}4]{2^n w_i}A_i^{(U_\XC)}$. From now on, we will drop the superscript ${U_\XC}$ of the Mercer basis for a clearer presentation (only in this section). We define a linear operator $\mathcal{N}$ describing the post-processing,
\begin{align}
    \mathcal{N}(O) = \sum_{i=1}^{4^n} \sqrt{2^n w_i} \tr(OA_i) A_i.
\end{align}
where $\mathcal{N}$ transforms $\rho({\bm x})$ into $\Tilde{\rho}({\bm x})$. The GTQK can then be written as
\begin{align}
    k(\vec{x},\vec{x'}) = \tr(\Tilde{\rho}(\vec{x})\Tilde{\rho}(\vec{x'})) = \tr(\mathcal{N}(\rho(\boldsymbol{x}))\mathcal{N}(\rho(\vec{x'}))).
\end{align}

With the operator $\mathcal{N}$ defined, we will show that $\phi_i (\vec{x}) = \frac{\tr(\rho(\vec{x}) A_i)}{\sqrt{\gamma_i}}$ are the eigenfunctions of $k$ with eigenvalues $2^n w_i \gamma_i$. We define the integral operator for $k$ as
\begin{align}
    (T_k f)(\boldsymbol{x}) = \int k(\boldsymbol{x},\vec{x'}) f(\vec{x'}) \mu(d \vec{x'}).
\end{align}
Now, we apply this integral operator to the eigenfunctions $\phi_j (\vec{x})$
\begin{align}
    (T_k \phi_j)(\boldsymbol{x}) &= \int k(\boldsymbol{x},\vec{x'}) \phi_j(\vec{x'}) \mu(d \vec{x'})\\
    &= \frac{1}{\sqrt{\gamma_j}}\tr( \int \rho(\vec{x'}) \otimes \mathcal{N}(\rho(\vec{x'})) \mu(d\vec{x'}) \cdot A_j \otimes \mathcal{N}(\rho(\boldsymbol{x}))) \\
    &= \frac{1}{\sqrt{\gamma_j}} \tr(L'(A_j) \mathcal{N}(\rho(\boldsymbol{x}))),
\end{align}
where $L'(A_j) = \tr_1(\int \rho(\vec{x'}) \otimes \mathcal{N}(\rho(\vec{x'})) \mu(d\vec{x'}) A_j^{(U_\XC)} \otimes \mathbb{1})$ and the last equality is obtained by repeating the trick in Appendix \ref{Appendix-Subsubsec:Eigendecomposition-GFQK}. As the operator $\mathcal{N}$ is linear, independent of $\vec{x'}$, and commutes with the $\tr_1$ operation, we can write $L'(A_j)$ in term of $L(A_j)$, i.e.: $L'(A_j) = \mathcal{N}(L(A_j))$. Hence,
\begin{align}
    L'(A_j) &= \mathcal{N}(L(A_j))\\
    &= \mathcal{N}\Big(\sum_i \gamma_i \tr(A_i A_j) A_i \Big)\\
    &= \sqrt{2^n w_j} \gamma_j A_j.
\end{align}
Therefore we have
\begin{align}
    (T_k \phi_j)(\boldsymbol{x}) &= \frac{1}{\sqrt{\gamma_j}}\tr(L'(A_j) \mathcal{N}(\rho(\boldsymbol{x})))\\
    &= \frac{1}{\sqrt{\gamma_j}}\sum_{i=1}^{4^n} \sqrt{2^n w_i} \sqrt{2^n w_j} \gamma_j \tr(\rho(\boldsymbol{x})A_i)\tr( A_i A_j)\\
    &= 2^n w_j \gamma_j \phi_j(\boldsymbol{x}).
\end{align}
We conclude that the GTQK shares the eigenfunctions $\phi_i$ of the GFQK with the eigenvalues re-scaled by $2^n w_i$. The verification can be further simplified as
\begin{align}
    (T_k \phi_j)(\boldsymbol{x}) &= \int k(\boldsymbol{x},\vec{x'}) \phi_j(\vec{x'}) \mu(d\vec{x'})\\
    &= \sum_{i=1}^{4^n} 2^n w_i \gamma_i \phi_i(\boldsymbol{x}) \int \phi_i(\vec{x'})\phi_j(\vec{x'}) \mu(d \vec{x'})\\
    &= \sum_{i=1}^{4^n} 2^n w_i \gamma_i \phi_i(\boldsymbol{x}) \delta_{ij}\\
    &= 2^n w_j \gamma_j \phi_j(\boldsymbol{x}) \:,
\end{align}
but we show here the first proof as it is more convincing and it stems from the definition of GTQK, instead of the self-proclaimed eigendecomposition of the kernel. Hence, the eigendecomposition form of the GTQK is
\begin{align} \label{Appendix-eq:GTQK-eigenform}
    k(\vec{x},\vec{x}') &= \sum_{i=1}^{4^n} 2^n w_i \gamma_i \phi_i(\vec{x})\phi_i(\vec{x'}).
\end{align}
Substituting $\phi_i (\vec{x}) = \frac{\tr(\rho(\vec{x}) A_i)} {\sqrt{\gamma_i}}$ into Eq.~\eqref{Appendix-eq:GTQK-eigenform} and cancels some terms yield
\begin{align} 
    k(\vec{x},\vec{x}') &= \sum_{i=1}^{4^n} 2^n w_i \tr(\rho(\vec{x})A_i) \tr(\rho(\vec{x'})A_i),
\end{align}
which is equal to Eq.~\eqref{Appendix-eq:GTQK-eigenform-expanded} with $A_i \rightarrow A_i^{(U_\XC)}$. The corresponding Mercer quantum feature feature map is defined as
\begin{align}
    \Psi^{(U_\XC)}(\vec{x}) = \left (\psi_1^{(U_\XC)}(\vec{x}), \cdots, \psi_{4^n}^{(U_\XC)}(\vec{x}) \right)^T.
\end{align}

\section{Reproducing kernel Hilbert space for generalized trace-induced quantum kernels}
\label{Appendix-Sec:RKHS-GTQKs}

In this Appendix we will show different ways of constructing the reproducing kernel Hilbert space (RKHS) for the generalized trace-induced quantum kernels (GTQKs), i.e.: via Moore-Aronsajn \cite{schuld2021supervised} and Mercer \cite{kubler2021inductive,heyraud2022noisy} constructions. The theory is generic, therefore, it can be applied to both quantum data and classical data.

\subsection{Moore-Aronszajn construction of RKHS for generalized trace-induced quantum kernels}
Recall that the generalized trace-induced quantum kernel is defined as
\begin{align}
    k(\boldsymbol{x},\vec{x'}) = \sum_{i=1}^{4^n} 2^n w_i \tr(\rho(\boldsymbol{x}) A_i^{(U_\XC)})\tr(\rho(\boldsymbol{x'}) A_i^{(U_\XC)})
\end{align}
with $\sum_{i=1}^{4^n} w_i^2 = 1$. Following the Moore-Aronsajn construction in Appendix~\ref{Appendix-Subsec:RKHS_MA-construction}, we construct the RKHS for the GTQKs as
\begin{align}
    \mathcal{F}_{k} = \overline{\Bigg \{ f: f(\cdot) = \sum_{i\in \mathbb{N}} \alpha_i k(\cdot,\boldsymbol{x}_i), ~\alpha_i \in \mathbb{R} ~\forall i, ~||f||^2_{\mathcal{F}_k} < \infty  \Bigg \}}.
\end{align}
Given two functions in $\mathcal{F}_{k}$, $f(\cdot) = \sum_i \alpha_i k(\cdot,\boldsymbol{x}_i)$ and $g(\cdot) = \sum_j \beta_j k(\cdot,\boldsymbol{x}_j)$, the inner product is defined as $\langle f,g \rangle_{\mathcal{F}_{k}} = \sum_{ij} \alpha_i \beta_j k(\boldsymbol{x}_i, \boldsymbol{x}_j)$ with $\alpha_i, \beta_j \in \mathbb{R}$. This relation establishes the reproducing property of the kernel, i.e.: $f(\boldsymbol{x}) = \langle f(\cdot), k(\cdot,\boldsymbol{x}) \rangle_{\mathcal{F}_{k}}$ $~\forall f \in \mathcal{F}_{k}$. Note that $k(\cdot,\boldsymbol{x})$ is the canonical feature map of the kernel as $\langle k(\cdot,\boldsymbol{x}), k(\cdot,\boldsymbol{x}') \rangle_{\mathcal{F}_{k}} = k(\boldsymbol{x},\boldsymbol{x}')$.

By the representer theorem, the quantum model can be written as
\begin{align} \label{Eqn:Quantum-model-GFQK}
    f(\boldsymbol{x},{\bm \alpha}) &= \sum_{i=1}^N \alpha_i \tr(\tilde{\rho}(\boldsymbol{x}_i)\tilde{\rho}(\boldsymbol{x})) =\tr ( \tilde{M}({\bm \alpha}) \tilde{\rho}(\boldsymbol{x})),
\end{align}
where $\tilde{M}({\bm \alpha}) = \sum_{i=1}^N \alpha_i \tilde{\rho}(\boldsymbol{x}_i) $ and $N$ is the number of training data. The optimal parameters ${\bf \alpha}^*$ would depend on the training data \cite{schuld2021supervised}. Therefore, $\mathcal{F}_{k_n}$ can be regarded as an alternative feature space to $\mathcal{F}$, and this feature map maps data to functions instead of matrices. 

\subsection{Mercer construction of RKHS for generalized trace-induced quantum kernels}
\label{Appendix-Subsec:Mercer-RKHS-GTQK}

The RKHS of the GTQK constructed using the Mercer quantum feature map $\Psi_{i}^{(U_\XC)}(\cdot)=\sqrt{2^n w_i } \tr(\rho(\cdot)A_i^{(U_\XC)})$ is defined as
\begin{align} \label{Appendix-eq:RKHS-QMercer}
    \mathcal{H}_{k}^{(U_\XC)} = \Bigg \{ f: f(\cdot) = \sum_{i=1}^{4^n} \tilde{\alpha}_i \sqrt{2^n w_i } \tr(\rho(\cdot)A_i^{(U_\XC)}), ~\tilde{\alpha}_i \in \mathbb{R} \Bigg\}.
\end{align}
Given two arbitrary functions $f(\cdot) = \sum_i \tilde{\alpha}_i \Psi_{i}^{(U_\XC)}(\cdot)$ and $g(\cdot) = \sum_j \tilde{\beta}_j \Psi_{j}^{(U_\XC)}(\cdot)$, the inner product in this space is defined as
\begin{align}
    \langle f, g \rangle_{\mathcal{H}_{k}^{(U_\XC)}}  &:= \sum_{i,j=1}^{4^n} \tilde{\alpha}_i \tilde{\beta}_j \left \langle \Psi_{i}^{(U_\XC)}(\cdot), \Psi_{j}^{(U_\XC)}(\cdot) \right \rangle_{\mathcal{H}_{k}^{(U_\XC)}} = \sum_{i=1}^{4^n} \tilde{\alpha}_i \tilde{\beta}_i,
\end{align}
and the kernel still has the reproducing property $f(\boldsymbol{x}) = \langle f(\cdot), k(\cdot, \boldsymbol{x}) \rangle_{\mathcal{H}_{k}^{(U_\XC)}}$ that enforces the orthogonality condition of $\Psi_i^{(U_\XC)} (\cdot)$, i.e.: $\left \langle \Psi_{i}^{(U_\XC)}(\cdot), \Psi_{j}^{(U_\XC)}(\cdot) \right \rangle_{\mathcal{H}_{k}^{(U_\XC)}} = \delta_{ij}$.

Alternatively, the RKHS for the GTQK can be constructed using the kernel eigenfunctions $\phi_{i}^{(U_\XC)}(\cdot)$ as a basis,
\begin{align} \label{Eqn:RKHS-GTQK-eigen}
    \mathcal{H'}_{k}^{(U_\XC)} = \left \{ f': f'(\cdot) = \sum_{i=1}^{4^n} \tilde{\alpha}_i \phi_{i}^{(U_\XC)}(\cdot), ~ \tilde{\alpha}_i \in \mathbb{R} \right\},
\end{align}
with the inner product between two arbitrary functions in $\mathcal{H'}_{k}^{(U_\XC)}$, $f' = \sum_i \tilde{\alpha}_i \phi_{i}^{(U_\XC)}(\cdot)$ and $g' = \sum_i \tilde{\beta}_i \phi_{i}^{(U_\XC)}(\cdot)$ defined as $\langle f', g' \rangle_{\mathcal{H'}_{k}^{(U_\XC)}}  := \sum_{i=1}^{4^n} \frac{\tilde{\alpha}_i \tilde{\beta}_i}{2^n w_i\gamma_{i}}$, and the squared norm of the function $f'$ is given by
\begin{align} \label{Eqn:square-norm-eigen}
    ||f'||^2_{\mathcal{H'}_{k}^{(U_\XC)}} &= \langle f', f'\rangle_{\mathcal{H'}_{k}^{(U_\XC)}} = \sum_{i=1}^{4^n} \frac{\tilde{\alpha}_i^{2}}{2^n w_i \gamma_{i}}.
\end{align}
The inductive bias of the quantum model can be understood from the optimization of the quantum models in the RKHS $\mathcal{H'}_{k}^{(U_\XC)}$. As shown in Eq.~\eqref{Eqn:RKHS-GTQK-eigen}, the quantum models could be written in term of the basis of the kernel eigenfunctions, i.e.: $f'(\boldsymbol{x}) = \sum_j \tilde{\alpha}_j \phi_j^{(U_\XC)}(\boldsymbol{x})$ with squared norm given by Eq.~\eqref{Eqn:square-norm-eigen}, transforming the cost function in Eq.~\eqref{Eqn:Cost-function} into 
\begin{align}
    \LC_{\vec{\tilde{\alpha}}}(S) = & \sum_{i=1}^{N} \ell (f'(\boldsymbol{x}_i),y_i) + \sum_{j=1}^{4^n} \frac{\tilde{\alpha}_j^{2}}{2^n w_j \gamma_j} \; .
\end{align}
While eigenvalues $2^n w_j \gamma_j$ do not appear in $f$, they are in the denominator of the regularization term in the cost function. In the regularization term, the coefficients $\tilde{\alpha}_j$ for the corresponding eigenfunctions are weighted by the inverse of the corresponding eigenvalues. The lower the eigenvalue, the more the corresponding eigenfunction is penalized. While one cannot change $\gamma_j$ after the quantum feature map is defined, one can always adjust the weights $w_j$ to tune the inductive bias of the model. Setting $w_j = 0$ implies infinite suppression of the corresponding eigenfunctions, effectively removing them from the optimization. Together with the other picture introduced in Sec.~\ref{Subsec:GTQK}, this provides a consistent understanding of how the inductive bias is being imposed by the weights $w_j$, i.e.: by re-scaling the eigenvalues and bias towards the eigenfunctions with larger weights. 

Now, we will show that $\mathcal{H}_{k}^{(U_\XC)}$ and $\mathcal{H'}_{k}^{(U_\XC)}$ are isometric isomorphic to each other. Since the integral operator $T_k$ shares its eigenfunctions and eigenvalues with the associated kernel $k$, therefore, the action of $T_k$ on an arbitrary function $f'(\cdot) = \sum_i \tilde{\alpha}_i \phi_{i}^{(U_\XC)}(\cdot)$ is just a re-scaling of parameters $\tilde{\alpha}_i$ by the corresponding eigenvalues
\begin{align}
    (T_{k}f')(\boldsymbol{x}) = \sum_{i=1}^{4^n} \tilde{\alpha}_i 2^n w_i\gamma_{i}\phi_{i}^{(U_\XC)}(\boldsymbol{x}).
\end{align}
The operator $T_{k}^{1/2}$ for which $T_{k} = T_{k}^{1/2} \circ T_{k}^{1/2}$ then re-scales parameters $\tilde{\alpha}_i$ by $\sqrt{2^n w_i\gamma_{i}}$. Hence, $T_{k}^{1/2}$ induces an isometric isomorphism between $\mathcal{H}_{k}^{(U_\XC)}$ and $\mathcal{H'}_{k}^{(U_\XC)}$. Note that the functions $f(\boldsymbol{x})$ in Eq.~\eqref{Appendix-eq:RKHS-QMercer} are actually the quantum neural network models with an additional scaling factor $\sqrt{2^n w_i}$
\begin{align}
    f(\boldsymbol{x}) = \sum_{i=1}^{4^n} \tilde{\alpha}_i \sqrt{2^n w_i} \tr(\rho(\boldsymbol{x}) A_i^{(U_\XC)})
\end{align}
expressed in the Mercer basis.

\end{document}